\documentclass{aastex}
\usepackage{emulateapj5}

\slugcomment{To appear in The Astrophysical Journal Supplement Series October 2003 issue}
\shorttitle{[OIII] images of Seyferts}
\shortauthors{Schmitt et al.}
\begin{document}

\title{A Hubble Space Telescope Survey of Extended [OIII]$\lambda$5007\AA\
Emission in a Far-Infrared Selected Sample of Seyfert Galaxies:
Observations\altaffilmark{1}}

\author{H. R. Schmitt\altaffilmark{2,3,4}, J. L. Donley\altaffilmark{5,6},
R. R. J. Antonucci\altaffilmark{7}, J. B. Hutchings\altaffilmark{8},
A. L. Kinney\altaffilmark{9}}

\altaffiltext{1}{Based on observations made with the NASA/ESA Hubble Space
Telescope, which is operated by the Association of Universities for Research
in Astronomy, Inc., under NASA contract NAS5-26555.}
\altaffiltext{2}{National Radio Astronomy Observatory, 520 Edgemont Road,
Charlottesville, VA22903.}
\altaffiltext{3}{Jansky Fellow.}
\altaffiltext{4}{email:hschmitt@nrao.edu}
\altaffiltext{5}{Department of Astronomy and Astrophysics, The Pennsylvania
State University, 525 Davey Laboratory, University Park, PA16802}
\altaffiltext{6}{National Radio Astronomy Observatory, P.O. Box 0, Socorro,
NM87801.}
\altaffiltext{7}{University of California, Santa Barbara, Physics Department,
Santa Barbara, CA 93106.}
\altaffiltext{8}{Dominion Astrophysical Observatory, Herzberg Institute of
Astrophysics, National Research Council of Canada, 5071 West Saanich Road,
Victoria, BC V9E 2E7, Canada.}
\altaffiltext{9} {NASA Headquarters, 300 E Street SW, Washington, DC20546.}

\begin{abstract}
We present a Hubble Space Telescope (HST) survey of extended [OIII] emission
for a sample of 60 Seyfert galaxies (22 Seyfert 1's and 38 Seyfert 2's),
selected based on their far infrared properties. The observations for 42 of
these galaxies were done in a snapshot survey with WFPC2. The remaining 18
were obtained from the HST archive, most of which were
observed with the same configuration. These observations cover
68\% of the objects in the sample defined by Kinney et al. (2000), and
create a valuable dataset for the study of the Narrow Line Region (NLR)
properties of Seyfert galaxies. In this paper, we present the details of
the observations, reductions, and measurements. We also discuss the 
extended structure of individual sources, and the relation of this emission 
to the radio and host galaxy morphology. We also address how representative 
the subsample of [O III]-imaged galaxies is of the entire sample, and 
possible selection effects that may affect this comparison of the properties 
of Seyfert 1 and Seyfert 2 galaxies.
\end{abstract}

\keywords{galaxies:active -- galaxies:Seyfert -- galaxies:structure --
galaxies:emission lines -- galaxies:nuclei -- surveys}

\section{Introduction}

The Unified Model for Seyfert galaxies (Antonucci 1993; Urry \& Padovani 1995)
is based in part on the idea that the nuclear region of these galaxies is
surrounded by a dusty torus. Under this scenario, the distinction between
Seyfert 1's and Seyfert 2's depends on whether the nucleus is observed through
the torus pole (Seyfert 1's) or equator (Seyfert 2's). Starting with the
spectropolarimetric observations of NGC\,1068 (Antonucci \& Miller 1985)
several tests have been applied to this model, which is now supported by
several lines of evidence, and is widely accepted by the AGN community
as broadly, if not universally, applicable.

Some of the most important evidence for this model are the observation of a
deficit of ionizing photons in Seyfert 2 galaxies (Neugebauer et al. 1980;
Wilson, Ward \& Haniff 1988; Storchi-Bergmann, Wilson \& Baldwin 1992),
and the collimated escape of radiation from the nucleus. The collimated
radiation is observed in the form of ionization cones, seen in the light of
high ionisation gas such as [OIII]$\lambda$5007\AA, or ionization maps created
by dividing [OIII] images by H$\alpha+$[NII]$\lambda$6548,84\AA\ ones
(Pogge 1988a,b,1989; Haniff, Wilson \& Ward 1988; Mulchaey, Wilson \&
Tsvetanov 1996a,b).

The most extensive study of the morphologies and sizes of the NLRs of Seyferts
was done by Mulchaey et al. (1996a), using ground-based observations. However,
due to the fact that in most of these galaxies the emission is concentrated
in the inner 2\arcsec\ region around the nucleus, most of their sources
revealed only halo-like [OIII] emission. Higher spatial resolution
images, like those achievable with HST, are necessary for a detailed study of
the NLR of these sources.

So far, however, most of the studies done using HST images (Schmitt \&
Kinney 1996; Capetti et al. 1996; Falcke, Wilson \& Simpson
1998, Ferruit et al. 2000) involved small samples of galaxies, and
probably suffered from major selection effects. In an attempt to
solve this sample problem we performed an HST snapshot survey to observe
the extended [OIII] emission in a large and well-defined sample of Seyfert
galaxies. We consider it very important that our sample was selected to be
as free as possible of orientation biases that invalidate comparisons
between two populations posited to be similar except in orientation.
Here we present the results of this snapshot survey, which
obtained data for 42 galaxies, as well as similar data for another
18 Seyfert galaxies, obtained from the HST archive. We discuss the sample
selection, observations, data reductions, and measurements. We also compare
the distances and luminosities of the observed galaxies with those of the 
entire Kinney et al (2000) sample, to verify that they are a representative
subsample, and can be used to draw robust statistics on the Unified Model.
Similarly, we compare the distances, luminosities and limits of detection
of extended [OIII] emission of the observed Seyfert 1's to Seyfert 2's, to
address possible selection effects.

Combining these [OIII] observations with the optical and radio ground-based
observations that we have for this sample, we show that we have created a
valuable dataset for the study of the Unified Model,
the interaction between the radio and line emission in the NLR, and the
ionization mechanisms for the gas. The comparison of the properties of the
NLR's of Seyfert 1's and Seyfert 2's will be presented in a separate paper.

\section{Sample}

One of the most important aspects of Unified Model studies is sample
selection. During recent years, several papers have presented results
which seem to contradict the model predictions, but most of these
can be explained as due to selection effects (see Schmitt et al. 2001a;
Ho \& Ulvestad 2001 for a more detailed discussion). In order to 
test the Unified Model, one has to select a sample based on isotropic
properties - those independent of the orientation of the torus to the
line of sight. This will insure against selecting different types of Seyferts
from different parts of the luminosity function (e.g. intrinsically brighter
Seyfert 2's), or any other orientation-dependent biases.

In our recent work, we have used the sample of Seyfert
galaxies extracted from the catalog of warm infrared sources of de Grijp
et al. (1987, 1992), which were selected  based on their 60$\mu$m
luminosities and warm 25/60$\mu$m colors. Our sample, described in Kinney et
al. (2000), includes all the Seyfert galaxies with z$<$0.031 in this
catalog, for a total of 88 sources (29 Seyfert 1's and 59 Seyfert 2's).
Notice that Kinney et al. (2000) combined all Seyfert 1 and 1.5 galaxies
in the Seyfert 1 group, and all the Seyfert 1.8, 1.9 and 2 in the
Seyfert 2 group. We have made an extensive study of the properties of this 
sample (Schmitt et al. 2001a), which currently includes ground-based B and I
images (Schmitt \& Kinney 2000), and high resolution (3.6cm VLA) radio
images (Schmitt et al. 2001b, S01b hereafter) for most of the galaxies. 

\section{Observations}

All the observations presented in this paper are from HST. Most
of these observations (45 of the 63 galaxies), are from 
snapshot proposal 8598 (P.I. Schmitt). Data for another 18 galaxies were
available in the HST archive, observed for the following programs:
3982 (P.I. Baldwin), 5140 (P.I. Macchetto), 5745 (P.I. Ford), 6332, 6419,
8240 (P.I. Wilson) and  8259 (P.I. Whittle). Although some of the archival
data have been presented elsewhere, we have re-reduced them to present here
in a homogeneous manner, and as a check of the reduction process.
Observations of 3 sources in our snapshot survey failed, so the
total number of galaxies with [OIII] images is 60, which corresponds to
$\approx$70\% of the full 60$\mu$m sample (22 Seyfert 1's and 38 Seyfert 2's).
Of the three observations which failed, two were due to guide star acquisition
problems (UGC\,10683\,B and IRAS\,11215-2806), and one to the misfortune that
the nucleus of the galaxy fell on bad CCD columns (Fairall\,49).

The majority of the galaxies in the sample were observed with the WFPC2
camera. The exceptions are NGC\,3393, which was observed with the WF/PC1
camera in PC mode (pre-COSTAR), and MRK\,3, which was observed with the FOC
camera (post-COSTAR). Details on the reduction of these two sources are given
in Schmitt \& Kinney (1996). In the case of the galaxies observed with WFPC2,
only 2 out of the 61 objects had radial velocities smaller than
$\sim1500$ km s$^{-1}$ (NGC\,1068 and NGC\,1386) and consequently their on-band
images were observed with the narrow band filter F502N in the PC camera. The
remaining 59 galaxies had to be observed with the Linear Ramp Filter.

The WFPC2 Linear Ramp Filter has a strip where the central
wavelength varies along the position, with a bandwidth of approximately
1.3\% of the central wavelength ($\sim$65\AA), so it can only be used
to image sources smaller than $\sim13^{\prime\prime}$. Previous ground-based
observations (Mulchaey et al. 1996a,b) show that most Seyfert galaxies have
NLR's with sizes smaller than $\sim2^{\prime\prime}$, indicating that a
$13^{\prime\prime}$ field of view is a good match for our project.

An important detail about observations with the Linear Ramp Filter
is that the coordinates of the sources have to be known with
an accuracy of $\sim1^{\prime\prime}$ or better, to place them in
the region of the filter which covers the wavelength of
interest. We checked this by comparing the position where the source 
was expected to fall on the CCD (calculated using the redshifted wavelength 
of the [OIII] line, and the information in the HST Manual), with the observed
position of the nucleus of the galaxy. None of the observations
were missed because of pointing problems. Figure 1 presents the difference
between the expected and observed positions of the galaxies. This Figure
shows that in most of the cases the galaxy falls within 2$^{\prime\prime}$ from
the expected position, with the median distance being 1.12$^{\prime\prime}$.
These displacements from the expected position
are consistent with the coordinate and pointing uncertainties.
The most deviant point in the sample corresponds to NGC\,4968, which falls
at 6.63$^{\prime\prime}$ from the expected position. This is mostly due to
the uncertainty in the coordinates of the galaxy (8$^{\prime\prime}$).
According to Pogge (1989), the ground based [OIII] image of this source
shows only an unresolved point source, so the large shift from the expected
position did not influence our results.

Another detail of the Linear Ramp Filter observations is the fact that,
for those galaxies with radial velocity larger than $\sim$8600~km~s$^{-1}$,
the redshifted [OIII]$\lambda$5007\AA\ line no longer falls on the second
WF camera chip, but actually on the PC camera. Although observations with the
PC camera would in some sense be desirable, because of the smaller pixel
size and better sampling, the sensitivity of this camera is significantly lower
than that of the WF camera. This would require a much larger integration
time to achieve sensitivities similar to the ones obtain with the WF camera,
thus reducing the likelyhood of the source being observed as a snapshot.
In cases like this, we took into account the fact that the bandwidth of the
Linear Ramp filter is 1.3\% of the central wavelength ($\sim4000$~km~s$^{-1}$)
and forced the galaxies with radial velocities between 8600~km~s$^{-1}$ and
9100~km~s$^{-1}$ to fall on the second WF camera chip, at the position
corresponding to [OIII]$\lambda$5007\AA\ redshifted by 8600~km~s$^{-1}$.
The only exception to
this procedure is NGC\,7674, which was extracted from the archive
(project 8259). This galaxy was observed with the Linear Ramp filter on the
PC camera. 

The on-band observations of our snapshot survey had total exposure times
between 600 and 1000 seconds, with the stronger sources having shorter
integrations and the fainter ones having longer integrations. The exposures
were split into 2 integrations of same duration, to allow the easier
elimination of cosmic rays. The observations taken from the archive were
obtained with similar total exposure times, and were also split into 2 or more
individual exposures. 

In order to be able to subtract the host galaxy contribution to the on-band
images, we obtained short exposure continuum images using the intermediate
band filter F547M. This filter is in a line-free region of the spectrum, and
the contribution from redshifted [OIII] emission is expected to be small,
because in all cases this line falls at wavelengths where the transmission is
less than 50\% of that of the peak of the filter. Due to time
constraints in our snapshot survey, we did not obtain multiple exposures
with this filter, but only a single 80-second integration per galaxy.
Most of the galaxies extracted from the archive also used this filter
to obtain continuum images, but usually had multiple exposures.
In the case of the project 6332, the continuum observations were also done
with the Linear Ramp Filter, instead of F547M.

Table 1 presents a summary of the observations, where we give the names of the
galaxies, their coordinates, exposure times, the names of the datasets in the
HST archive, the proposal which observed the galaxy, and comments. Each entry
in this Table is divided into 2 or 3 lines, with the first ones corresponding
to the on-band observations, and the last one corresponding to the
continuum observations.

\section{Reductions and measurements}

The data reduction for the continuum images taken with filter F547M, and
the on-band images taken with the filter F502N, followed the standard HST
pipeline procedures. In the case of the observations taken with the
Linear Ramp Filter, most of the reductions were done by the HST pipeline,
with the exception of the flat field correction. Since there are no flat
fields for this filter, we used the one for the filter F502N, which has a
similar wavelength and bandwidth.

The on-band and continuum images were aligned using either a star in the same
chip as the galaxy, or the nucleus of the galaxy when stars were not
available. Multiple images were combined to eliminate cosmic rays. For
those galaxies with only one continuum image, cosmic rays were first eliminated
using an automatic routine, and the remaining ones eliminated by hand.
All combined and cleaned on-band and continuum images were further inspected
for cosmetic defects and cleaned.

Some extra care was necessary with the reduction of archive observations
obtained by the project 6332, 8240 and 8259. Project 6332 used the Linear Ramp
Filter to observe the continuum images, which resulted in the on-band and
continuum images being located in different WF chips. A similar problem
happened with the data taken from project 8240 and one of the galaxies
observed by 8259 (NGC\,4507), where the continuum observations were taken
with the PC camera and the on-band images were taken with the WF camera.
In these cases the four WFPC2 chips were first mosaiced, to correct for
distortions and put the images on the same scale, and then aligned and
combined to eliminate cosmic rays.

The flux calibration of the continuum images taken with filter F547M, and the
on-band images taken with filter F502N was done in the standard way, using the
information available on the image headers. The calibration of the images
taken with the Linear Ramp Filter was done with the WFPC2 Exposure Time
Calculator for extended sources. For a source with a given flux, observed for
a given exposure time at the redshifted [OIII]$\lambda$5007\AA\ wavelength,
one is able to use the Exposure Time Calculator to determine 
its count rate and obtain a calibration coefficient. Based
on these numbers, we estimate the photometric calibration for the different
wavelengths observed has an accuracy of $\sim$5\%.

The background for both on-band and continuum images was defined by averaging
the mean counts in several regions around the galaxy. This value was
subtracted from the corresponding images. Continuum free [OIII] images were
created by subtracting a scaled continuum image from the on-band image. The
scaling of the continuum images was estimated from the bandpass FWHM of 
the on-band filter. In some cases it was necessary to change this
value by a small amount, to avoid over- or under-subtraction. Finally, we
used emission free regions of the continuum-free [OIII] images to determine
their r.m.s. ($\sigma$), cliped these images at 3$\sigma$ above the
background level and rotated them to put N up and E to the left.

The continuum-free [OIII] images were used to measure the [OIII] flux of
these galaxies. We measured the nuclear fluxes, obtained using a circular
aperture of a 100 parsecs in radius centered at the nucleus, and the total
[OIII] fluxes, obtained integrating the flux inside a rectangular box placed
around the regions of emission.

We present in Table 2 the names and alternate names of the galaxies, their
Seyfert types, radial velocities, and the 3$\sigma$ surface brightness
detection limit of their [OIII] images, which also corresponds to the
first contour level of the Figures. This Table also presents the [OIII]
fluxes obtained integrating the flux above the 3$\sigma$ detection limit
(F([OIII])$_{int}$), the nuclear [OIII] fluxes (F([OIII])$_{nuc}$),
integrated inside a circular aperture with radius of 100 parsecs centered
at the nucleus, the [OIII] fluxes obtained from the literature
(F([OIII])$_{lit}$), the integrated [OIII] luminosities (calculated
assuming H$_0=75$~km~s$^{-1}$~Mpc$^{-1}$, which will be used throughout
this paper) and the reference from which F([OIII])$_{lit}$ was obtained.

As a check on our flux calibration, we present in Figure 2 a
comparison between the observed fluxes and the ones obtained from the
literature. This Figure shows a good agreement between the two measurements,
with most of the points lying close to the bisector line, or to the right
of it. This indicates that on average our images get larger fluxes than the
published ones - a result which is expected, since the ground based data were
obtained from long-slit observations which would miss the most extended
regions of emission. The average difference between the observed values and
the published ones is 0.09$\pm$0.25~dex for all the galaxies, 0.17$\pm$0.26~dex
for Seyfert 1's and 0.05$\pm$0.23~dex for Seyfert 2's.

To address how representative the observed sample is in relation to the entire
60$\mu$m sample, and to investigate possible biases influencing our analysis,
we compare the properties of the 60 galaxies with observed [OIII] images to
those of the galaxies in the 60$\mu$m sample which were not observed,
as well as the properties of Seyfert 1's and Seyfert 2's with observed [OIII]
images. We do not compare the observed sample to the entire 60$\mu$m sample,
because all the values for that sample are repeated in this one, and such a
comparison would be biased. Figure 3a (left) compares the distances of the
galaxies in the 60$\mu$m sample and the observed sample, which show very
similar distributions. The KS test shows, from
the comparison between the observed sample and the 
galaxies which were not observed, that there is an 84.6\% probability that
two samples drawn from the same parent population would differ this much.
This result shows that the observations were not biased towards galaxies
closer or farther away from us. The comparison between the distances of
Seyfert 1's and Seyfert 2's with observed [OIII] images (Figure 3b) gives
a similar result, with the KS showing a 81.6\% probability that two
samples drawn from the same parent population would differ this much or more.

It is also important to verify if there are any selection biases in the
luminosities of the observed galaxies, or in the limit of detection
of extended emission in Seyfert 1's and Seyfert 2's. The comparison between
the [OIII] luminosities of the observed sources and the entire 60$\mu$m sample,
where we used [OIII] fluxes from the literature for the sources which were not
observed by HST, is presented in Figure 4a. We can see in this Figure that
there is a good agreement between the two samples. Comparing the luminosities
of the observed galaxies with those of the galaxies which were not observed,
the KS test gives 38.6\% probability that two samples
drawn from the same parent population would differ this much. In the case
of Seyfert 1's and Seyfert 2's with observed [OIII] images, their
distributions are shown in Figure 4b. We can see in this Figure that the two
Seyfert types have similar distributions, with the KS test
giving a 6.4\% probability that two samples
drawn from the same parent population would differ this much.

One of the major objectives of this survey is to compare the sizes of the NLR's
of Seyfert 1's and Seyfert 2's. This requires us to show that there is no bias
towards the detection of more extended emission in one type of Seyfert
relative to the other. We do this by comparing the 3$\sigma$ detection limit of
the observed Seyfert 1's and Seyfert 2's. Since most of the images were obtained
with the same instrument and with similar exposure times, most galaxies have
very similar $3\sigma$ surface brightness detection limits, with a distribution
peaked around 3.7$\times10^{-17}$~erg~cm$^{-2}$~s$^{-1}$~pix$^{-1}$.
The KS test shows that there is no statistically significant
difference in the distribution of detection limit fluxes of Seyfert 1's and
Seyfert 2's, with a probability of 35.4\%
that two sample drawn from the same parent population would differ as much as
these two. However, since the distance of our galaxies vary by a factor of 10,
a more important test consists of comparing the luminosities of the detection
limits. We find that in our case they also are very similar, with the KS
test showing a 96.4\% probability that two samples
drawn from the same parent population would differ this much or more.

The final [OIII] images of the galaxies are presented in Figures 5 through 14
(6 galaxies per Figure). The lowest contour value of these images corresponds
to the 3$\sigma$ level above the background in surface brightness, and the
following ones increase in powers of 2 times 3$\sigma$. We also indicate in
these images the position of the nucleus, measured on the continuum images,
the position angle of the host galaxy major axis, measured by Schmitt \&
Kinney (2000) on ground based images, and the position angle of the radio
emission published in Kinney et al. (2000) and Schmitt et al. (2001).
With the exception of MRK\,348, we used radio jet position angles determined
based on VLA measurements, because these observations correspond
to dimensions similar to those sampled by the [OIII] images. In those cases
where the jet changes direction (e.g. MCG+8-11-11, NGC\,1068 and MRK\,3),
we plot the position angle value measured closer to the nucleus.

These [OIII] images were used to measure the extension of the emission in
these galaxies. We measured the effective radius, extent of the photometric
semi-major and semi-minor axes of the [OIII] emission, R$_e$, R$_{Maj}$ and
R$_{Min}$, respectively, as well as the position angle (PA) of the [OIII]
major axis, and the offset between the galaxy nucleus and the center of the
[OIII] emitting region. The effective radii were determined by integrating
the [OIII] flux inside circular apertures, centered at the nucleus, with
radii increasing in steps of 0.1 pixels. The aperture which contains half of
the integrated [OIII] flux gives the value of R$_e$. The measurements of
R$_{Maj}$, R$_{Min}$ and the PA of the [OIII] emission were done directly
by eye on the images, using as reference the 3$\sigma$ contour levels. The
measurement of the offset between the nucleus and the center of the [OIII]
emission was done in the following way. Based on the [OIII] images, it is
possible to divide the major axis of emission into two segments, one to each
side of the nucleus. We call these segments X1 and X2, which correspond to the
larger and smaller segment, respectively. The offset between the nucleus
and the center of the [OIII] emitting region is calculated using the
expression: Offset=(X1-R$_{Maj}$)/R$_{Maj}$, which gives the displacement
in units of the semi-major axis of emission.

Table 3 gives the measured values of R$_e$, R$_{Maj}$, R$_{Min}$, PA,
Offset from the nucleus to the [OIII] emission center, and the
Figure number for the [OIII] image. Some of the galaxies in the sample
have conically-shaped NLR's, however, their R$_{Maj}$ is not always along
the axis of the apparent cone. In these cases, Table 3 also gives
the PA of the cone axis in parentheses in column 5.

\section{Individual Objects}

In this section, we describe the observed emission of individual galaxies and
give a brief review of some data in the literature.
For each galaxy, we present the sizes of the major and minor [O III] emission
axes, the PA of the major axis, discuss the structure of the NLR (e.g. 
apparent cone shape, or blobs of emission) and point out
relations of these structures to the radio jet and host galaxy. It is not our
intent to do a detailed study of the data available in the literature for each
one of the galaxies. However, we make an effort to point out which galaxies
have published radio images, as well as ground-based and HST narrow band
images. Also, for the Seyfert 2's, we indicate which ones are known to show
polarized broad emission lines.

The galaxies in this section are ordered by Right Ascension (Table 2).  Notice
that the dimensions given in this section correspond to the total extent
of emission along the major and minor axis (based on the 3$\sigma$ surface
brightness contour), and not only the size of the 
semi major and minor axes (R$_{Maj}$ and R$_{Min}$) given in Table 3.

\subsection{MRK\,348}          

This is a Seyfert 2 galaxy with polarized broad emission lines (Tran 1995).
The HST [OIII] and H$\alpha+$[NII] images were studied by Capetti et al.
(1996) and Falcke, Wilson \& Simpson (1998). The [OIII] emission (Figure 5 top
left) is extended by 2.85$^{\prime\prime}$ (840 pc) along P.A.$=185^{\circ}$,
with a blob of emission at approximately 0.9$^{\prime\prime}$ S of the
nucleus. Along the perpendicular direction the emission is extended by
1.75\arcsec\ (510 pc). Close to the nucleus, the emission is extended along
P.A.$\sim-10^{\circ}$, which is similar to the direction of the radio emission
observed by Ulvestad et al. (1999). It is possible to draw a cone with opening
angle 60$^{\circ}$ on top of the extended [OIII] emission.

\subsection{MRK\,573}          

This is a Seyfert 2 galaxy known from ground-based images to have extended
[OIII] emission (Haniff et al. 1988; Pogge et al. 1993, 1995) aligned with
the radio triple source observed by Ulvestad \& Wilson (1984), and misaligned
by 60$^{\circ}$ from the host galaxy major axis. The HST images of this galaxy
were studied in detail by SK96, Capetti et al. (1996), Falcke et al. (1998)
and Ferruit et al. (1999). Figure 5 (top right) shows that the emission is
composed of several arches around the nucleus, with a total extent of
8.9$^{\prime\prime}$ (2980 pc) along P.A.$=120^{\circ}$, and 3.7\arcsec\
(1250 pc) along the perpendicular direction. A cone with opening angle of
75$^{\circ}$ can be drawn on top of the [OIII] emission. 

\subsection{IRAS\,01475-0740}  

This is a Seyfert 2 galaxy with strong radio emission (S01b; Thean et al.
2001), unresolved on VLA scales, and also known to have polarized broad
emission lines (Tran 2001). The [OIII] emission (Figure 5; middle left)
has a halo-like morphology with an average extent of 0.8$^{\prime\prime}$
(275pc).

\subsection{ESO\,153-G\,20}  

The [OIII] image of this Seyfert 2 galaxy is presented in Figure 5 (middle
right), where we can see that the bulk of the line emission comes from the
inner 1$^{\prime\prime}$ and is extended along P.A.$=-100^{\circ}$, misaligned
about 70$^{\circ}$ from the host galaxy major axis. Some diffuse emission is
seen extending over a region of $\sim1.85\times$2.65\arcsec (710$\times$1010
pc) centered at the nucleus, with a major-axis at PA$=-10^{\circ}$. A cone
with opening angle of 80$^{\circ}$ can be drawn on top of the extended
emission close to the nucleus.

\subsection{MRK\,590}        

This Seyfert 1 galaxy has a slightly extended NLR (Figure 5, bottom left), with
dimensions of 1.1$\times$1.5\arcsec (560$\times$770 pc), and major axis along
PA$=-5^{\circ}$. This emission does not resemble a conically-shaped NLR.
Most of the extended [OIII] emission comes from the region North of the nucleus
and is misaligned by $\sim50^{\circ}$ relative to the host galaxy major axis.
The radio emission (S01b) is unresolved.

\subsection{MRK\,1040}       

The [OIII] image of this Seyfert 1 galaxy is presented in the bottom right
panel of Figure 5. The [OIII] emission is extended and shaped like a cone with
opening angle of $120^{\circ}$, apex at $\sim0.6$\arcsec\ NW of the nucleus,
and the axis pointing towards the SE direction, perpendicular to the host
galaxy disk. The emission along the apparent cone axis is extended by
1.1\arcsec\ (350 pc), and by 1.5\arcsec\ (480 pc) along the perpendicular
direction. The radio emission of this galaxy is unresolved (S01b).

\subsection{UGC\,2024}         

This Seyfert 2 galaxy presents diffuse [OIII] emission with no clear conical
shape. The emission is extended by 1.45$\times$2.6\arcsec\ (630$\times$1130 pc)
with the major axis along P.A.$=-35^{\circ}$, within 5$^{\circ}$ from the host
galaxy major axis (Figure 6, top left). The emission along the host galaxy
minor axis is extended by 1.45\arcsec\ (630 pc). This galaxy present only
an unresolved nuclear radio source (S01b).

\subsection{NGC\,1068}         

The [OIII] image of this Seyfert 2 galaxy is shown in the top right panel of
Figure 6. This image shows that most of the [OIII] emission originates in a
V-shaped region with opening angle of 50$^{\circ}$, extending for approximately
10\arcsec\ (750pc) along PA$=35^{\circ}$, and 5.8\arcsec\ (430 pc) in the
direction perpendicular to the cone. The direction of
this emission is coincident with the
radio jet observed by Wilson \& Ulvestad (1982b). Emission extending to regions
farther away from the nucleus can be seen towards the NE and SE. 
NGC1068 is the most extensively studied Seyfert 2 galaxy in the literature.
It was the first one where polarized broad emission lines were detected
(Antonucci \& Miller 1985), as well as the first one to be shown to
have a conically shaped NLR (Pogge 1988a). The HST images of this galaxy were
discussed in detail by Evans et al. (1991), Macchetto et al. (1994),
Capetti, Axon \& Macchetto (1997), Bruhweiler et al. (2001). Studies of the
kinematics of the NLR gas show outflows and a strong interaction between the
gas and the radio source (Axon et al. 1998, Cecil et al. 2002). The radio
emission is extended along  PA$=33^{\circ}$ on the larger scales (Wilson \&
Ulvestad 1982b), but is aligned along the N-S direction in the nuclear region
(Muxlow et al. 1996; Gallimore et al. 1996). The high resolution X-ray images
obtained with Chandra (Young, Wilson
\& Shopbell 2001) also show extended emission aligned the radio and [OIII].

\subsection{MRK\,1058}        

This is a Seyfert 2 galaxy with faint and unresolved radio emission (S01b).
The [OIII] image (Figure 6, middle left) shows that the line emission is
extended by 1.4$\times$2.3\arcsec\ (470$\times$ 760 pc) along
PA$=205^{\circ}$. This emission has a V-shaped morphology with opening
angle of 55$^{\circ}$, and is close to perpendicular to the
host galaxy major axis. The peak of emission is concentrated around the
nucleus with a second blob at 0.75\arcsec\ SW.

\subsection{MCG\,-02-08-039} 

The middle left panel of Figure 6 presents the [OIII] image of this Seyfert 2
galaxy, which shows only diffuse emission around the nucleus, with a diameter
of $\sim$2\arcsec\ (1160pc). The radio emission is unresolved (S01b).

\subsection{UGC\,2514}       

This is a Seyfert 1 galaxy with slightly extended radio emission along
PA$=236^{\circ}$ (S01b). The [OIII] image (Figure 6, bottom left panel)
shows that most of the line emission is displaced to the W of the nucleus,
with two blobs separated by $\sim$0.6\arcsec\ along PA$=-70^{\circ}$. The
emission is extended by 1.35\arcsec\ (350 pc) along this direction, and by
1.15\arcsec (290 pc) along P.A.$=190^{\circ}$.

\subsection{IRAS\,03106-0254}  

Figure 6 (bottom right) shows the [OIII] image of this Seyfert 2 galaxy.
The emission is extended by $0.9\times$1.5\arcsec\ (470$\times$790 pc) along
PA$=30^{\circ}$, which is at a direction similar to that of the extended
radio emission (S01b), and misaligned by 45$^{\circ}$ from the host
galaxy major axis.

\subsection{IRAS\,03125+0119}  

The [OIII] emission of this Seyfert 2 galaxy (Figure 7, top left)
is diffuse, with the largest extension of 1.5\arcsec\ (700 pc) along the
host galaxy major axis, and 1\arcsec\ (470 pc) along the minor axis.
The radio emission is unresolved (S01b).

\subsection{MRK\,607}          

The top right panel of Figure 7 shows that the [OIII] emission of this
Seyfert 2 galaxy is extended by 3.75\arcsec\ (660 pc) along the host galaxy
major axis, and 1.35\arcsec\ (240 pc) along the minor axis.
Most of the emission originates at the nucleus, or in a blob
between 0.9-1\arcsec\ NW of the nucleus. The radio emission is unresolved
(Nagar et al. 1999).  Ferruit et al. (2000) present a detailed discussion of
the HST images of this galaxy. 

\subsection{ESO\,116-G\,18}  

The [OIII] image of this Seyfert 2 galaxy (Figure 7, middle left) is extended
by 1.7\arcsec\ (610 pc) along PA$=75^{\circ}$, and 1\arcsec\ (360 pc)
in the perpendicular direction. Most of the emission originates from
regions N of the nucleus.

\subsection{NGC\,1386}         

This is a Seyfert 2 galaxy known, based on observations from the ground, to
have extended [OIII] and radio emission (Weaver, Wilson \& Baldwin 1991;
Storchi-Bergmann et al. 1996). The image presented in Figure 7 (middle right
panel) shows that the [OIII] emission consists of several blobs along the N-S
direction, misaligned by 25$^{\circ}$ from the host galaxy major axis,
and extending over a region of 5.9\arcsec\ (330 pc). The emission along the
perpendicular direction is extended by only 1.6\arcsec (90 pc).
There are 3 blobs of emission to the N of the nucleus, located at 0.8\arcsec,
1.3\arcsec and 2.8\arcsec, while to the S there is a blob at 1\arcsec\ and
a ring of blobs at 2\arcsec\ from the nucleus.
The direction of the extended [OIII] emission is similar to the extended
radio emission detected by Nagar et al. (1999).  Ferruit et al. (2000) present
a detailed study of this and other HST narrow band images of this galaxy.

\subsection{ESO\,33-G\,02}   

The [OIII] image of this Seyfert 2 galaxy is presented in the bottom left
panel of Figure 7. The emission is distributed in a more or less homogeneous
way around the nucleus, extended by 1$\times$1.45\arcsec\ (350$\times$510 pc),
with the major-axis along the N-S direction, perpendicular to the host galaxy
major-axis. Some diffuse emission can be seen to the NE of the nucleus.
The ground based [OIII] images published by Mulchaey et al. (1996a) show
only slightly resolved emission.

\subsection{MCG\,-05-13-017} 

This Seyfert 1 galaxy presents [OIII] emission homogeneously distributed
around the nucleus (Figure 7 bottom right). This emission is elongated along
the host galaxy major axis, extending for approximately 2\arcsec\ (500 pc)
in this direction and by 1.5\arcsec\ (380 pc) along the minor axis.
This galaxy has only unresolved nuclear radio emission (S01b).

\subsection{MCG\,+08-11-011} 

Previous HST images obtained with WF/PC1 did not show any extended [OIII]
emission (Bower et al. 1994; SK96) which was probably due to the strong
spherical aberration of the pre-COSTAR images. Radio images from S01b show
3.6cm emission extended for 3.5\arcsec\ along the N-S direction. Closer to the
nucleus Ulvestad \& Wilson (1986) showed that the radio structure consists of
a triple source extended by $\sim$1\arcsec\ along P.A.$=127^{\circ}$.
The top left panel of Figure 8 presents the [OIII] image of this Seyfert 1
galaxy. We can see that most of the emission is extended close to the N-S
direction, along P.A.$=20^{\circ}$, aligned with the large-scale radio emission.
The total extent of the emission in this direction is 3.6\arcsec\ (1450 pc),
and 3\arcsec\ (1210 pc) along the perpendicular direction. South of the
nucleus we see a CCD defect which could not be removed from the data,
but also does not influence the results. A large portion of the emission
is concentrated at the nucleus with a second blob of emission located at
0.7\arcsec\ N of it. Closer to the nucleus, we see two
blobs of emission, one to each side of it, separated by
$\sim$1\arcsec\ along P.A.$=130^{\circ}$. These structures may be related to
the nuclear triple radio source detected by Ulvestad \& Wilson (1986).

\subsection{MRK\,3}            

The [OIII] image of this Seyfert 2 galaxy is S-shaped, with a clear conical
NLR at regions close to the nucleus (Figure 8; top right). The major extent
of the line emission is 2.2\arcsec\ (580 pc) along the E-W direction, while
along the N-S direction it is extended by 1\arcsec\ (260 pc). The line
emission is aligned with the radio emission described by Ulvestad \& Wilson
(1984). A detailed discussion of this galaxy is given by SK96
and Capetti et al. (1996). Capetti et al. (1999) and Ruiz et al. (2001)
discuss the kinematics of the NLR gas, which is strongly influenced by the
interaction with the radio emission.

\subsection{UGC\,3478}         

The middle left panel of Figure 8 shows that the [OIII] emission of this
Seyfert 1 galaxy is diffuse, extended by 1.15$\times$1.65\arcsec\
(290$\times$410 pc), with the major axis along PA$=55^{\circ}$.
This emission is aligned close to the host galaxy major axis. The radio
emission shows only an unresolved source (S01b).

\subsection{MRK\,6}          

Ground-based observations of this Seyfert 1 galaxy (Haniff et al. 1988) showed
only an unresolved nuclear source. Our observations (Figure 8, middle right)
show [OIII] emission extended by 3.2\arcsec\ (1150 pc) along P.A.$=-10^{\circ}$,
a direction similar to that of the extended radio emission observed by S01b.
The emission is concentrated at the nucleus, with a series of blobs around it
and a finger of emission extending for 0.5\arcsec\ towards P.A.$=210^{\circ}$.
The largest blob is located 0.7\arcsec\ N of the nucleus, and is probably
related to the emission detected in the radio. Another small blob is observed
at 0.5\arcsec\ NW of the nucleus and is possibly related to the begining
of a transverse radio structure observed by Kukula et al. (1996) and S01b.

\subsection{Fairall\,265}

The [OIII] image of this Seyfert 1 galaxy is presented in the bottom left panel
of Figure 8. The emission is diffuse and extended by 2.9$\times$3.45\arcsec\
(1660$\times$1970 pc) along PA$=-20^{\circ}$.

\subsection{MRK\,79}         

This is a Seyfert 1 galaxy with extended radio emission along
PA$=11^{\circ}$ (Ulvestad \& Wilson 1984; S01b; Thean et al. 2000),
also known from ground-based images to have extended [OIII] emission
(Haniff et al. 1988). Our [OIII] image (Figure 8 bottom right) is extended by
4.6\arcsec\ (1980 pc) along the N-S direction, which appears closely related
to the radio emission. To the S of the nucleus we see two blobs of emission,
one at 0.6\arcsec\ and another at 1.2\arcsec\ from the nucleus. A larger number
of structures is seen to the N, where we detect blobs at 0.5\arcsec,
0.9\arcsec\ and 1.6\arcsec. The northern tip of the  emission is separated by
2.5\arcsec\ from the nucleus and is composed of several blobs along a
3.25\arcsec\ region in the E-W direction.

\subsection{MRK\,622}        

Nagar et al. (1999) detected slightly resolved radio emission in this Seyfert 2
galaxy, extended along PA$=0^{\circ}$. Mulchaey et al. (1996a) found only
halo like [OIII] emission in ground-based observations. Our [OIII] image
(Figure 9 top left) is extended by 0.95$\times$1.3\arcsec\
(430$\times$580 pc) along P.A.$=55^{\circ}$, perpendicular to the host
galaxy major axis.

\subsection{ESO\,18-G\,09}   

This Seyfert 2 galaxy has an apparent V-shaped [OIII] emission, with an opening
angle of 45$^{\circ}$, extended by 3\arcsec\ (1040 pc) along PA$=-30^{\circ}$
(Figure 9 top right). One blob of emission is seen at 0.7\arcsec\ NW of the
nucleus and another one at the same distance towards the SE. Emission in
the direction perpendicular to the apparent cone axis is extended by 1.1\arcsec\
(380 pc).

\subsection{MCG\,-01-24-012} 

The middle left panel of Figure 9 shows the [OIII] image of this Seyfert 2
galaxy. The emission is extended for 1.15$\times$2.3\arcsec\
(440$\times$880 pc) with the major axis along PA$=75^{\circ}$. Most of the
emission is concentrated around the nucleus and in a blob at 0.5\arcsec\ W
of it. The radio emission is extended in the E-W direction (S01b).

\subsection{MRK\,705}        

Ground-based narrow band [OIII] images of this Narrow Line Seyfert 1 galaxy
(Mulchaey et al. 1996a), as well as radio images (S01b), showed only an
unresolved source. Our HST [OIII] image (Figure 9 middle right) is compact,
extending for a diameter of 1.1\arcsec\ (620 pc) around the nucleus.

\subsection{NGC\,3281}         

This Seyfert 2 galaxy was studied in detailed by Storchi-Bergmann, Wilson \&
Baldwin (1992), who used ground-based narrow band imaging and long-slit
spectroscopy. These authors detected a cone of ionized gas extending up to
2~Kpc above the disk of the galaxy, and an outflow along the axis of this cone.
The radio emission studied by S01b does not show any extended emission.
Our [OIII] image (Figure 9 bottom left) confirms the ground-based [OIII]
observations of Storchi-Bergmann et al. (1992), showing a conically-shaped
NLR with opening angle $\sim80^{\circ}$ towards the NE. The emission is
extended by 3.9\arcsec\ (800pc) along the cone axis, and 6.1\arcsec\ (1260 pc)
along the N-S direction. The emission inside the cone
can be divided into two major structures, one consisting of several blobs
close to the nucleus, and the other one being a ridge of emission 2\arcsec\ NE
of the nucleus, parallel to the host galaxy major axis. We also detect a 
small blob of emission at 4.5\arcsec\ S of the nucleus, corresponding to the
counter ionization cone, which is mostly hidden by the host galaxy disk.
The emission observed by HST seems to be less extended than that observed
by Storchi-Bergmann et al. (1992), however, this can be due to the field of
view of the Liner Ramp Filter.

\subsection{NGC\,3393}         

The [OIII] emission of this Seyfert 2 galaxy (Figure 9 bottom right)
is S-shaped, extended by 5.3\arcsec\ (1410 pc) along PA$=65^{\circ}$
and 2.8\arcsec\ (740 pc) along the perpendicular direction.
The S structure surrounds the radio emission observed by S01b.
A detailed discussion of the NLR of this galaxy, involving the observations
presented here, ground based imaging and spectroscopy, and radio continuum
observations, is presented by Cooke et al. (2000). This galaxy is known to
have polarized broad emission lines (Kay, Tran \& Magalh\~aes 2002).

\subsection{UGC\,6100}         

Figure 10, top left panel, presents the [OIII] image of this Seyfert 2 galaxy.
The emission is S-shaped and extended by 1.3$\times$3.5\arcsec\
(740$\times$1990 pc) along the N-S direction. The radio image presented by
Kukula et al. (1995) is slightly resolved.

\subsection{NGC\,3516}         

The [OIII] image of this Seyfert 1 galaxy is presented in the top right panel
of Figure 10. The emission is S-shaped, extended by 13.6\arcsec\ (2330 pc)
along PA$=20^{\circ}$. The extension of the emission along the E-W direction
at the bottom and top of the S-shaped structure is 6.9\arcsec
(1180 pc). A detailed study of the NLR of this source is
presented by Ferruit et al. (1998). Nagar et al. (1999) detected extended
radio emission along PA$=10^{\circ}$.

\subsection{NGC\,3783}         

This Seyfert 1 galaxy has a halo like NLR with [OIII] emission extended over a
region of 1.9\arcsec\ (310 pc) in diameter around the nucleus (Figure 10 middle
left). Ground based narrow band observations by Winge et al. (1992) found only
an unresolved source, and the radio emission also is unresolved (S01a).

\subsection{MRK\,766}        

The [OIII] emission of this Seyfert 1 galaxy is concentrated around the
nucleus, as can be seen in the middle right panel of Figure 10. It has a
halo-like morphology, with a diameter of 1.9\arcsec\ (470 pc). The radio
emission is slightly extended along P.A.$=32^{\circ}$ (Nagar et al. 1999).

\subsection{NGC\,4388}         

This Seyfert 2 was one of the first galaxies where a conically shaped NLR was
detected (Pogge et al. 1988b; Yoshida et al. 2002). Galactic scale outflows
have been detected by Veilleux et al. (1999), and Matt et al. (1994) detected
soft X-ray emission extended over 4.5 kpc in observations with ROSAT. The
observations presented here, as well as radio observations, are discussed in
detail by Falcke et al. (1998). This galaxy is also known to have polarized
broad emission lines (Young et al. 1996). The [OIII] image (Figure 10 bottom
left) shows a V-shaped NLR with opening angle of 90$^{\circ}$ towards the
S, extended over 6.1\arcsec\ (1000 pc) in this direction. The extent along the
perpendicular direction is 7.8\arcsec\ (1270 pc). Most of the emission comes
from regions S of the nucleus, except for some emission corresponding to the
counter cone, which is observed to the NE side of the nucleus. Most of the
counter cone emission is obscured by the host galaxy.

\subsection{NGC\,4507}         

The [OIII] image of this Seyfert 2 galaxy is presented in the bottom right
panel of Figure 10. We can see that the emission is elongated along
PA$=-35^{\circ}$ in the inner $\sim$2\arcsec\ region, becoming more circular
in the outer regions. The major extent of the emission is 3.5\arcsec\ (800 pc).
The bulk of the emission is related to the nucleus and a blob located at
1\arcsec\ to the NW. This galaxy show broad emission lines in polarized
light (Moran et al. 2000).

\subsection{NGC\,4593}         

The [OIII] image of this Seyfert 1 galaxy (Figure 11 top left) has a halo-like
morphology, slightly elongated along PA$=100^{\circ}$, which is a
direction similar to the host galaxy major axis. The extent of the emission in
this direction is 1.7\arcsec\ (300 pc), while along the minor axis the emission
is extended by 1.35\arcsec\ (240 pc). The radio emission
of this galaxy is unresolved (S01b).

\subsection{NGC\,4968}       

Figure 11, top right panel presents the [OIII] image of this Seyfert 2 galaxy.
The major axis of the [OIII] emitting region is extended by 2.2\arcsec\
(420 pc) along PA$=40^{\circ}$ and 1.3\arcsec\ (250 pc), along the
perpendicular direction. Previous ground-based observation
did not detect any extended [OIII] emission (Pogge 1989), and found only some
marginally extended radio emission (S01b). A more detailed description of the
NLR of this galaxy is given by Ferruit et al. (2000).

\subsection{IRAS\,13059-2407}  

The [OIII] emission of this Seyfert 2 galaxy is faint, mostly displaced to
the W relative to the peak of continuum emission, and extended by
0.5$\times$1.35\arcsec\ (140$\times$360 pc) along PA$=-10^{\circ}$ (Figure 11
middle left). The emission in this direction is divided into two blobs,
separated by 0.35\arcsec. The radio emission of this galaxy is unresolved
(S01b).

\subsection{MCG\,-6-30-15}     

The middle right panel of Figure 11 presents the [OIII] image of this Seyfert 1
galaxy. The emission is extended by 1.5$\times$3.9\arcsec\ (230$\times$590 pc),
with the major axis of the [OIII] emission at P.A.$=-65^{\circ}$. This
direction is at 10$^{\circ}$ from the host galaxy major
axis. The radio emission of this galaxy is unresolved (Nagar et al. 1999),
and a detailed description of the NLR emission is given by Ferruit et al.
(2000).

\subsection{NGC\,5347}         

Ground-based narrow-band images  of this Seyfert 2 galaxy (Pogge 1989;
Gonz\'alez Delgado \& P\'erez 1996) show a double structure along
PA$=25^{\circ}$. This galaxy is also known to have polarized broad
emission lines (Moran et al. 2001). The nuclear radio emission is unresolved
(S01b). Our [OIII] image (Figure 11 bottom left) confirms the two structures
detected on ground based images, a nuclear V-shaped structure with opening
angle $\sim60^{\circ}$, extending by 1.75\arcsec\ (260 pc) along
PA$=30^{\circ}$, and the second structure at 2.4\arcsec\ (370 pc) NE from
the nucleus, which has a width of 0.35\arcsec\ (50 pc) along the radial
direction from the nucleus and a length of 2.45\arcsec\ (370 pc).

\subsection{IRAS\,14082+1347}  

All the [OIII] in this Seyfert 2 galaxy (Figure 11 bottom right)
originates in a region $\sim$1\arcsec\ S of the nucleus (PA$=165^{\circ}$).
It has a major extent of 1\arcsec\ (310 pc) in the N-S direction and
0.75\arcsec\ (230 pc) in the E-W direction. The radio emission of this
galaxy is unresolved (S01b).

\subsection{NGC\,5548}         

The [OIII] emission from this Seyfert 1 galaxy has a halo-like morphology
(Figure 12 top left), with a diameter of 1.3\arcsec\ (450 pc) around the nucleus.
It also presents some low surface brightness fingers of emission extending by
2.7\arcsec\ (900 pc) in the N-S direction. The radio emission extends for
18\arcsec\ along PA$=-15^{\circ}$ (Wilson \& Ulvestad 1982a). Ground-based
observations (Wilson et al. 1989) detected low intensity [OIII] emission
which is not detected by the current observations. This could be due to
differences in the sensitivity of the two observations.

\subsection{IRAS\,14317-3237}  

Figure 12, top right panel presents the [OIII] image of this Seyfert 2 galaxy,
which is elongated by 1.4\arcsec\ (690 pc) along PA$=-5^{\circ}$, a direction
similar to the observed extended radio emission (S01b). The emission along the
perpendicular direction, close to the host galaxy minor axis, is extended by
0.75\arcsec\ (370 pc).

\subsection{UGC\,9826}         

The [OIII] emission of this Seyfert 1 galaxy (Figure 12 middle left) has a
halo morphology with a slightly larger extent of 1.55\arcsec\ (870 pc)
along the host galaxy major axis, and 0.9\arcsec\ (510 pc) along the minor
axis. The radio emission is unresolved (S01b).

\subsection{UGC\,9944}         

This Seyfert 2 galaxy has a conically shaped NLR aligned along 
PA$=90^{\circ}$,with an opening angle of 110$^{\circ}$
(Figure 12 middle right). The extent of
the NLR in this direction is 1.5\arcsec\ (710 pc), while in the perpendicular
direction it is extended by 2\arcsec\ (950 pc). The radio emission of this
galaxy consists of a triple source extended by approximately 8\arcsec\ along
PA$=67^{\circ}$. We do not detect any [OIII] emission related to the
outermost radio emission.

\subsection{IRAS\,16288+3929}  

The bottom left panel of Figure 12 presents the [OIII] image of this Seyfert 2
galaxy. The emission is extended by 1.6$\times$2.7\arcsec\ (940$\times$1590 pc)
along PA$=65^{\circ}$. The radio emission is unresolved (S01b).

\subsection{IRAS\,16382-0613}  

The [OIII] emission of this Seyfert 2 galaxy (Figure 12 bottom right) is
elongated along PA$=15^{\circ}$, extended by 1.2\arcsec\ (650 pc) in this
direction and 0.9\arcsec\ (480 pc) in the perpendicular direction.
The radio emission, observed by S01b, is unresolved.

\subsection{UGC\,10889}      

The top left panel of Figure 13 presents the [OIII] image of this Seyfert 2
galaxy. The largest extent of the emission is 1.25\arcsec\ (680 pc) along
PA$=95^{\circ}$ and consists of two blobs, one centered at the nucleus
and one 0.4\arcsec\ W of it. The radio emission is unresolved (S01b).

\subsection{MCG\,+03-45-003} 

The [OIII] emission of this Seyfert 2 galaxy (Figure 13 top right) is extended
by 3\arcsec\ (1410 pc) along PA$=15^{\circ}$, and 1.3\arcsec (600 pc) in the
perpendicular direction. The emission is distributed
among several blobs, one centered at the nucleus, one 0.75\arcsec\ S, one
0.35\arcsec\ N and another one 1\arcsec\ N. The radio emission of this galaxy
was studied by S01b, which found only an unresolved nuclear source.

\subsection{Fairall\,51}     

Figure 13, middle left panel shows the [OIII] image of this Seyfert 1 galaxy.
The emission is extended by 2.65\arcsec\ (730 pc) along the host galaxy major
axis. Most of the emission is concentrated at the 1\arcsec\ region around the
nucleus, with a finger like structure extending to the S. The overall emission
can be fit by an apparent cone with opening angle of 110$^{\circ}$ and the
axis along the SW direction.

\subsection{NGC\,6860}  

Ground-based narrow-band images of this Seyfert 1 galaxy, published by
L\'{\i}pari, Tsvetanov \& Macchetto (1993), show emission extended by 10\arcsec\
along the E-W direction. Our [OIII] image (Figure 13 middle right) shows
emission along the same direction, consisting of an S-shaped NLR along
PA$=85^{\circ}$, extended by $\sim$4.5\arcsec\ in this direction (1300 pc),
and 2.4\arcsec\ (690 pc) in the perpendicular direction. Most of the emission
is concentrated at the nucleus and in a blob 0.4\arcsec\ E, with a second more
extended and fainter structure at 0.9\arcsec\ E. To the W of the nucleus we
see a long arm of emission extending by $\sim$0.8\arcsec\ and a more diffuse
structure at 2.5\arcsec.

\subsection{UGC\,11630}      

Most of the [OIII] emission of this Seyfert 2 galaxy (Figure 13 bottom left) is
displaced to the E of the nucleus. It has a major extent of 1.6\arcsec\
(380 pc) along PA$=30^{\circ}$ and is divided into two blobs separated by
0.6\arcsec. The radio emission of this galaxy is unresolved (S01b).

\subsection{IC\,5063 = PKS\,2048-57}

The ground-based [OIII] image of this Seyfert 2 galaxy shows emission extended
over a region larger than $\sim$30\arcsec, along P.A.$=-65^{\circ}$ (Morganti
et al. 1998; Morganti et al. 1999). The line emission is aligned with the radio
emission and the host galaxy major-axis. This galaxy is also known to have
polarized broad emission lines (Inglis et al. 1993). Our [OIII] image confirms
the ground based results, but shows only a much smaller portion of the extended
emission, due to the field of view of the Linear Ramp Filter (Figure 13 bottom
right). This emission can be represented by a bicone centered at the nucleus,
with opening angle of 60$^{\circ}$, extending by 12\arcsec\ (2640 pc) along
PA$=-65^{\circ}$ and $\sim$3\arcsec\ (660 pc) along the perpendicular
direction. At distances larger than 3\arcsec\ N of the nucleus the emission
seems to split on the two sides of the host galaxy major axis, giving it the
appearance of an X-shaped NLR.

\subsection{NGC\,7212}         

This is a Seyfert 2 galaxy in an interacting system (Schmitt \& Kinney 2000),
which presents polarized broad emission lines (Tran 1995).
The [OIII] emission of this Seyfert 2 galaxy (Figure 14 top left)
is extended along PA$=170^{\circ}$, with dimensions of 2.1\arcsec\
by 4.8\arcsec (1080$\times$2480 pc). The emission is composed of several
individual blobs to the N and S of the nucleus. This emission is aligned
with the radio emission. A detailed study of the narrow band and radio
images of this galaxy is presented by (Falcke et al. 1998).

\subsection{NGC\,7213}         

The [OIII] emission of this Seyfert 1 galaxy (Figure 14 top right) has a halo
like morphology, homogeneously distributed around the nucleus, with a diameter
of 1.1\arcsec\ (130 pc). The radio emission is unresolved (S01b, Thean et al.
2000). The kinematics and the chemical abundance of the gas in this galaxy was
studied by Storchi-Bergmann et al. (1996).

\subsection{MRK\,915}        

Figure 14, middle left panel, presents the [OIII] image of this Seyfert 1
galaxy. The emission is very irregular, with a major extent of 4.1\arcsec\
(1910 pc) along PA$=5^{\circ}$, and 2.6\arcsec (1220 pc) in the perpendicular
direction. Most of the emission comes from the nucleus and a structure
0.65\arcsec\ to the SW. This structure has 1.3\arcsec\ in length
(PA$=140^{\circ}$) and 0.4\arcsec\ in width. Another blob of emission is
seen at 1.1\arcsec\ to the N of the nucleus. The radio emission, observed by
S01b, shows an unresolved nuclear structure and a faint tail of emission
extending to the SW.

\subsection{UGC\,12138}       

The [OIII] emission of this Seyfert 1 galaxy (Figure 14 middle right) has a
diffuse morphology with a diameter of 1.5\arcsec\ (710 pc) centered at the
nucleus. The radio emission observed by Kukula et al. (1995) is unresolved.

\subsection{UGC\,12348}       

Figure 14, bottom left panel, presents the [OIII] image of this Seyfert 2
galaxy, which is elongated along PA$=100^{\circ}$ with an extent of
1.15\arcsec$\times$2.2\arcsec (560$\times$1080 pc).
The radio emission, observed by S01b, is unresolved.

\subsection{NGC\,7674}         

This Seyfert 2 galaxy was observed to have polarized broad emission lines 
by Miller \& Goodrich (1990). The radio emission was studied by Unger et al.
(1988), Neff \& Hutchings (1992) and Kukula et al. (1995), who found a double
source separated by $\sim$0.5\arcsec\ along PA$=117^{\circ}$ and a smaller
component to the E of the  nucleus. Ground-based spectroscopy shows young
stars around the nucleus of this galaxy (Gonz\'alez Delgado, Heckman \&
Leitherer 2001). The HST [OIII] image, presented in the bottom right panel of
Figure 14, is extended by 4.4\arcsec\ (2480 pc) along PA$=120^{\circ}$,
which is a direction similar to that of the extended radio emission, and 
2.7\arcsec\ (1520 pc) along the perpendicular direction. One can also see in
this image, at 1.3\arcsec\ to the NW of the nucleus, along the host galaxy
major axis, a small loop and a hole in the emission. 

\section{Summary}

We have presented the results of an imaging survey of extended [OIII] emission
in a sample of 60 Seyfert galaxies (22 Seyfert 1's and 38 Seyfert 2').
We discussed the details of the different observations used in this
paper, the reduction and calibration techniques used to create the 
continuum-subtracted [OIII] images, and the measurements done with these 
images. By comparing the redshifts and [OIII] luminosities of the
observed galaxies to those of the entire 60$\mu$m sample (Kinney et al. 2000),
we find a good agreement, indicating that the observed galaxies are a
representative subsample and can be used to draw statistical results.
Furthermore, a comparison between the distances, [OIII] luminosities and
detection limits of the Seyfert 1's and Seyfert 2's with observed [OIII]
images shows a good agreement between these two groups of galaxies, which
suggests that selection effects will not influence the results of studies
which use these data to compare their properies.  This paper also gives a
description of the structure of the NLR of these galaxies, together with a
brief review of the relevant data available in the literature.

\acknowledgements

We would like to thank J. S. Ulvestad and J. E. Pringle for comments.
This work was partially supported by the NASA grants HST-GO-8598.07-A and
AR-8383.01-97A. The National Radio Astronomy Observatory is a facility of the
National Science Foundation, operated under cooperative agreement by Associated
Universities, Inc. This research made use of the NASA/IPAC Extragalactic
Database (NED), which is operated by the Jet Propulsion Laboratory, Caltech,
under contract with NASA. J.D.'s work was supported by the National Science
Foundation, through its Research Experience for Undergraduates program.

\clearpage

\begin{figure}
\epsscale{0.5}
\plotone{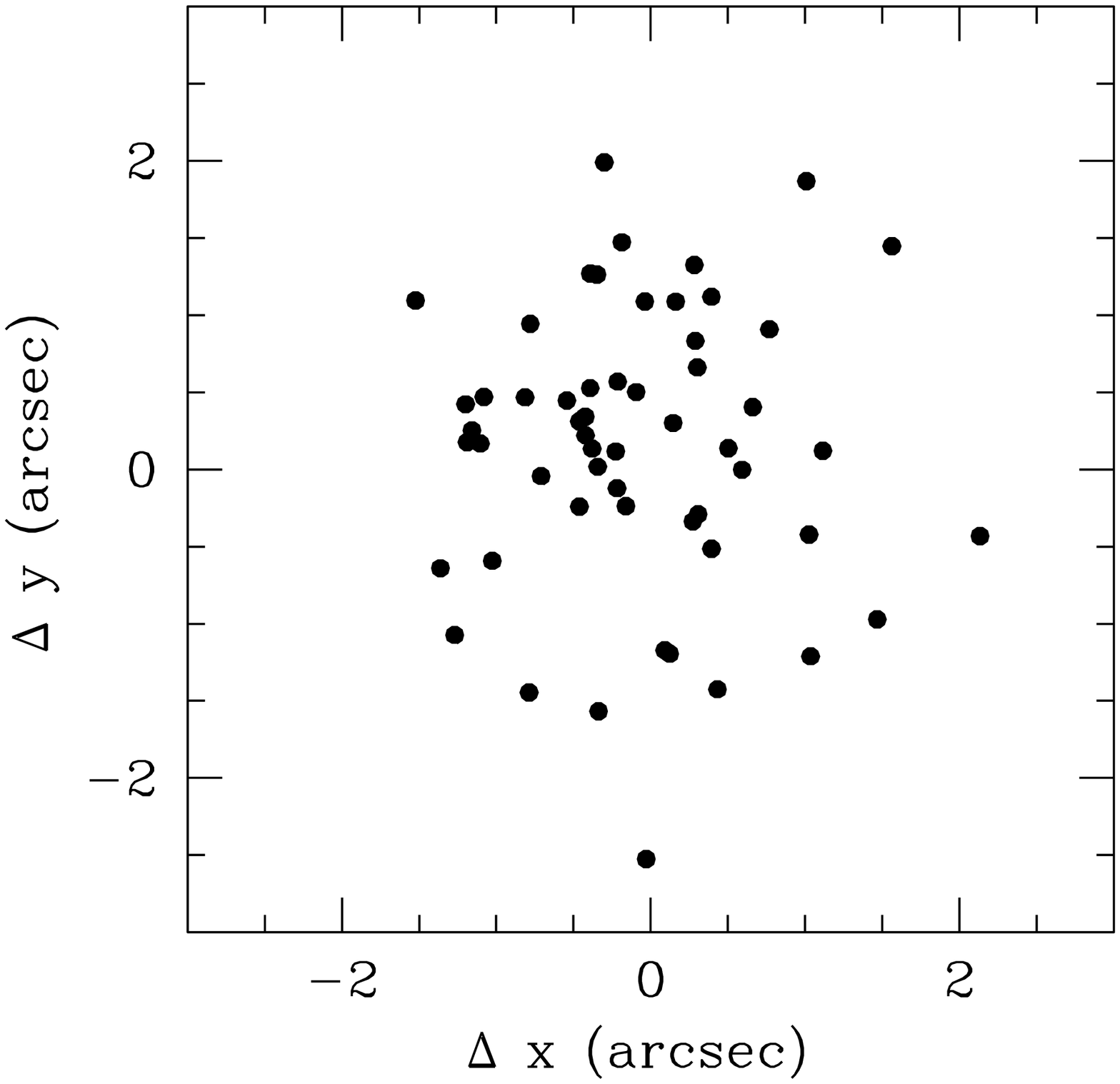}
\caption{Difference between the expected and observed position of the
source in the detector. This plot excludes the deviant point of NGC\,4968 at 
$\Delta$X$=-5.41^{\prime\prime}$ and $\Delta$Y$=-3.84^{\prime\prime}$.}
\end{figure}

\begin{figure}
\plotone{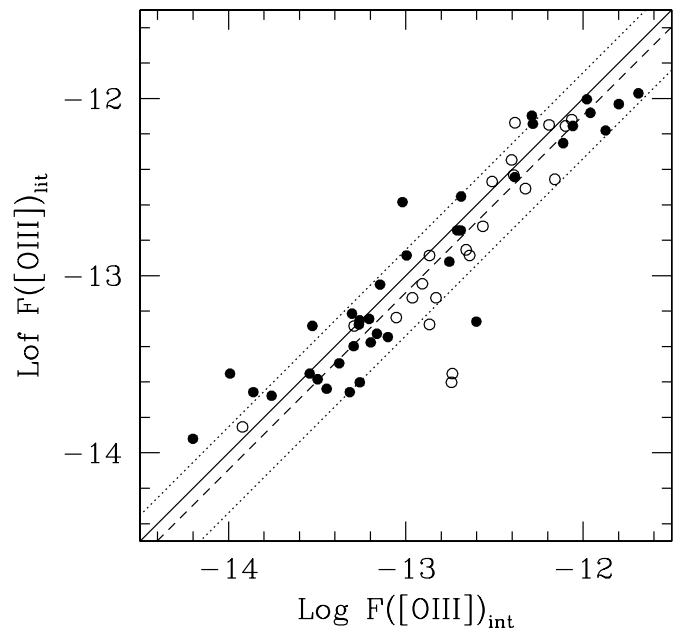}
\epsscale{0.5}
\caption{Comparison between the logarithm of the observed integrated [OIII]
fluxes (F([OIII])$_{int}$) and the logarithm of the [OIII] fluxes obtained
from the literature (F([OIII])$_{lit}$). Open symbols represent Seyfert 1's
and filled ones Seyfert 2's. The solid line is the bisector line. The dashed
line represents the average difference between the observed and published
[OIII] fluxes, and the dotted lines represent the 1$\sigma$ deviation.}
\end{figure}

\begin{figure}
\epsscale{1}
\plottwo{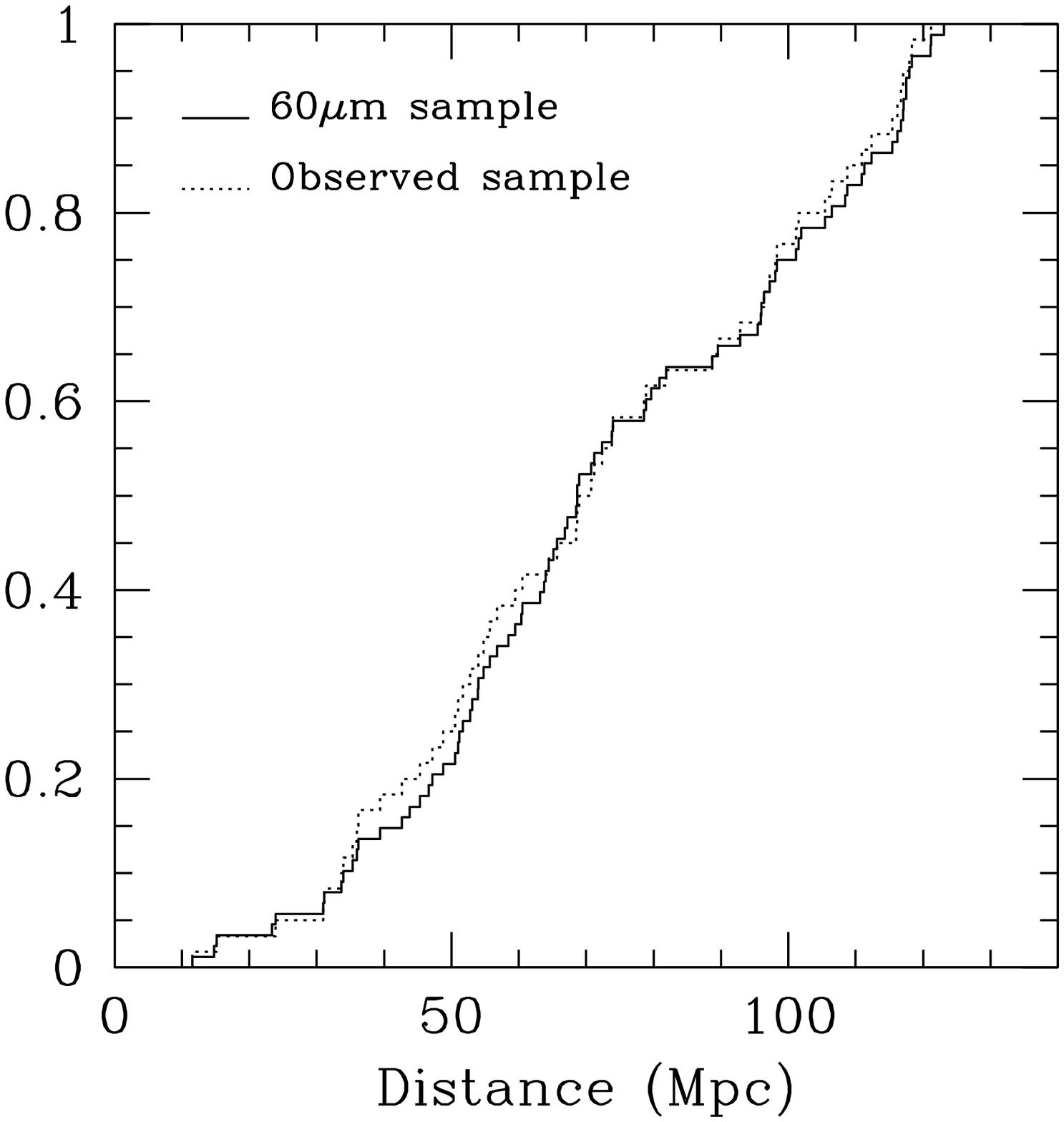}{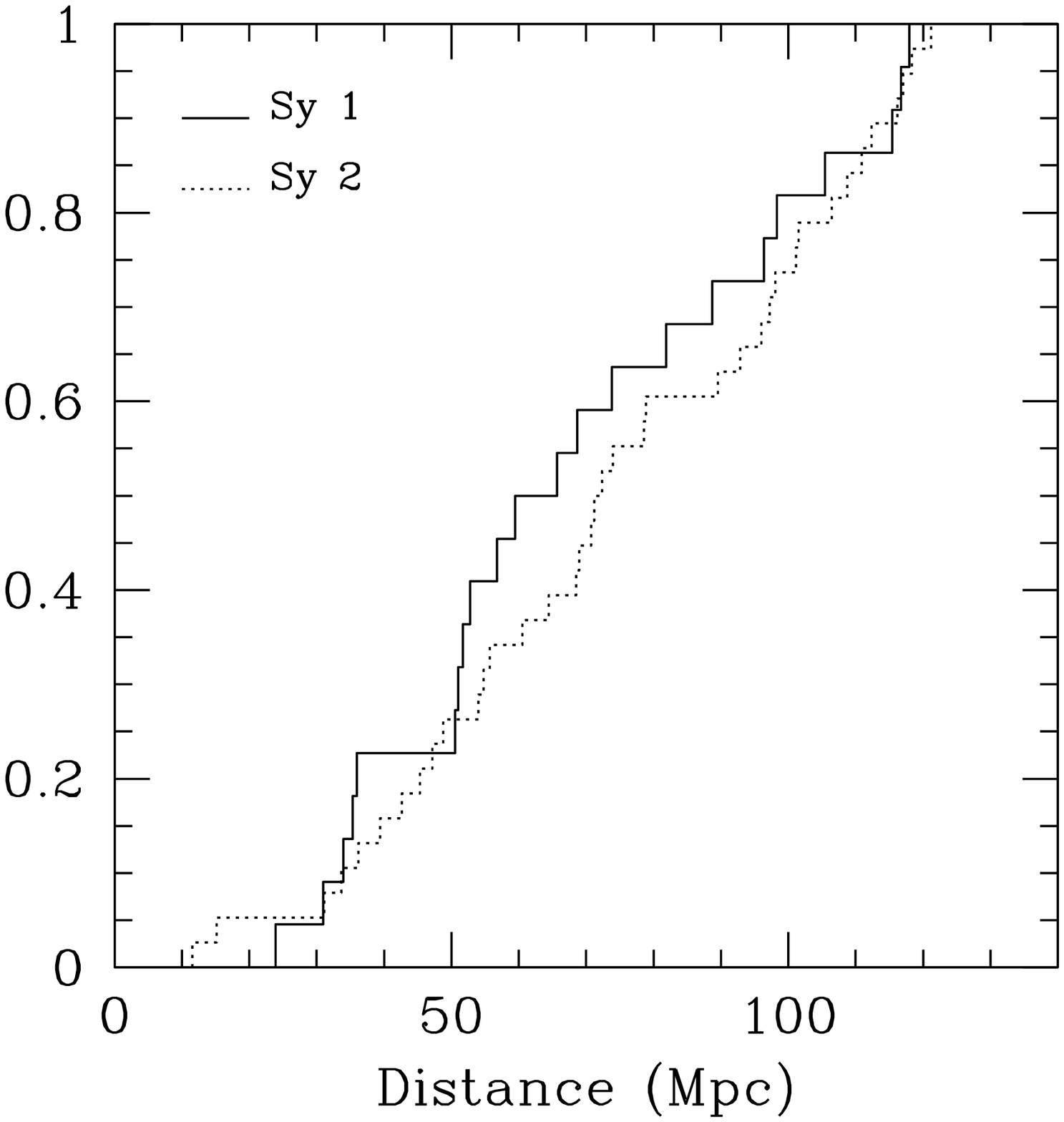}
\caption{Left: The cumulative distribution of distances for the galaxies
with [OIII] images (dotted line), and the 60$\mu$m sample (solid line).
Right: The cumulative distribution of distances for the Seyfert 2
(dotted line) and Seyfert 1 galaxies (solid line) with [OIII] images.}
\end{figure}

\begin{figure}
\plottwo{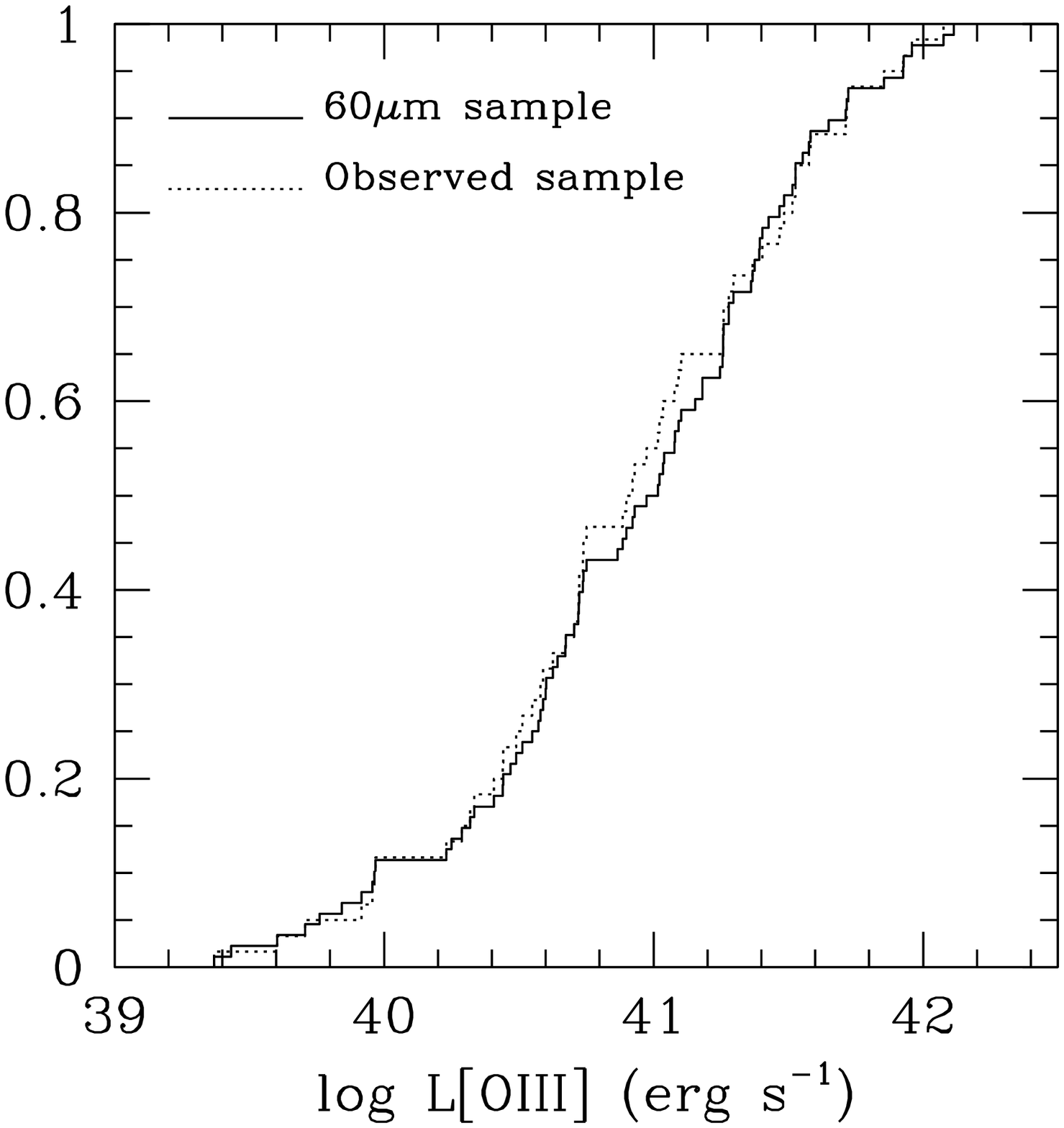}{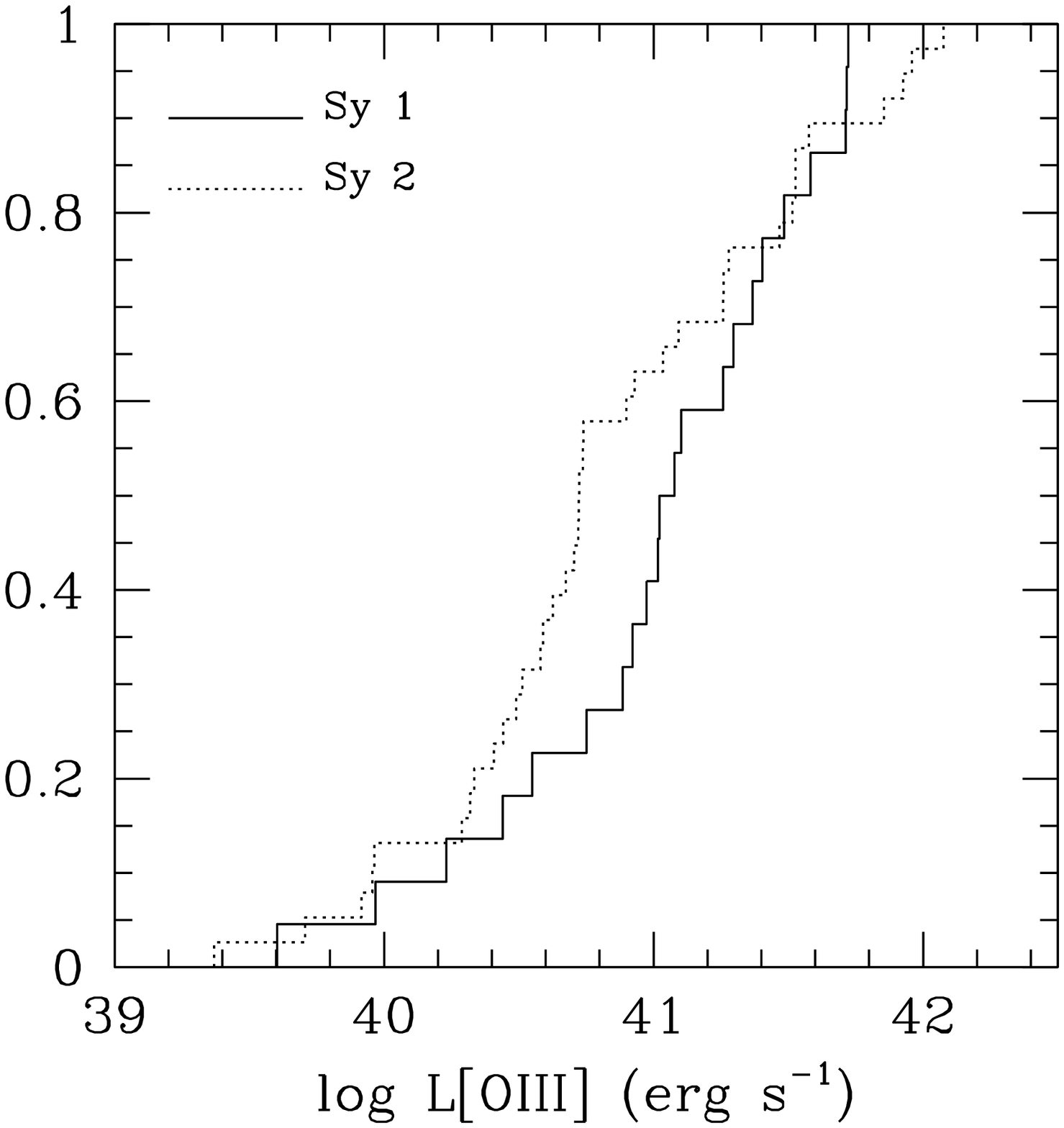}
\caption{Same as Figure 3, but for the logarithm of the integrated [OIII]
luminosities.}
\end{figure}

\vskip -2.0cm
\begin{figure}
\epsscale{0.8}
\plotone{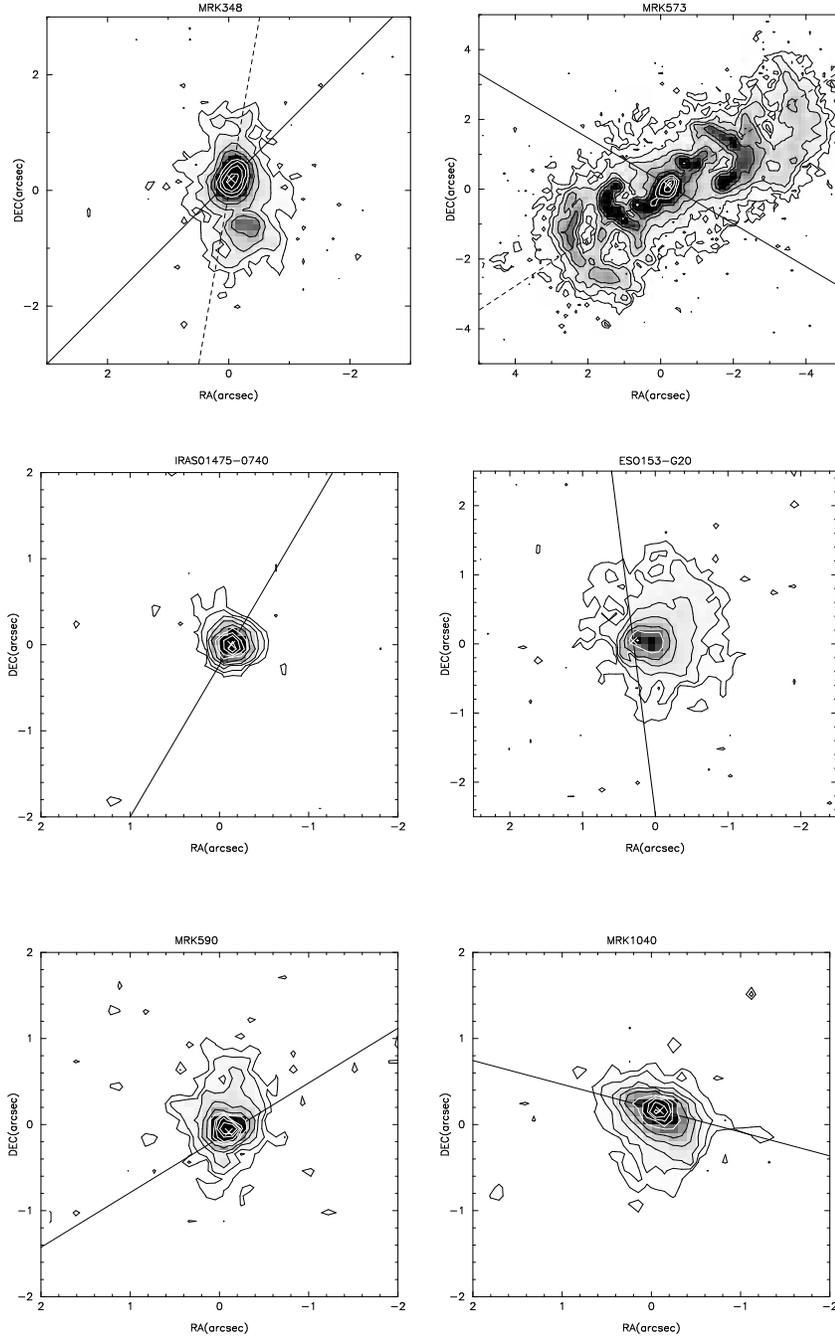}
\vskip -2.0cm
\caption{[OIII] continuum-subtracted image of MRK\,348, MRK\,573,
IRAS\,01475-0740, ESO\,153-G\,20, MRK\,590, MRK\,1040. The contours start
at the 3$\sigma$ level above the background and increase in powers of 2 times
3$\sigma$ (3$\sigma\times2^n$). The position of the nucleus, measured in
the continuum images, is plotted as a white cross. The position-angle of the
host galaxy major axis is shown as a solid line crossing the figure through
the nucleus. For those galaxies with extended radio emission, we present the
direction of the jet as a dashed line. North is up and East to the left.}
\end{figure}

\vskip -2.0cm
\begin{figure}
\epsscale{0.8}
\plotone{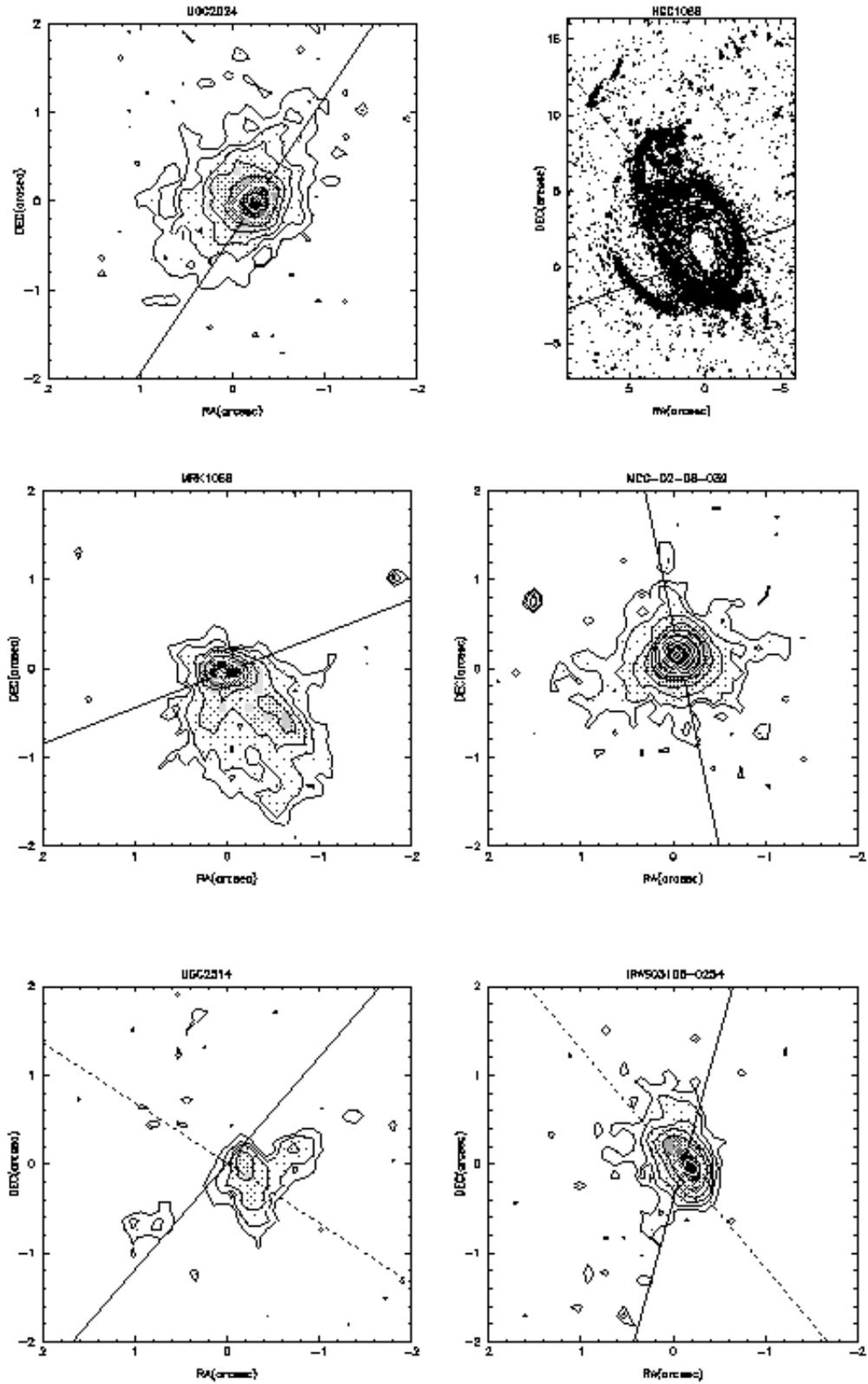}
\vskip -0.05cm
\caption{Same as Figure 5 for UGC\,2024, NGC\,1068, MRK1058, MCG\,-02-08-039, 
UGC\,2514 and IRAS\,03106-0254.}
\end{figure}

\begin{figure}
\plotone{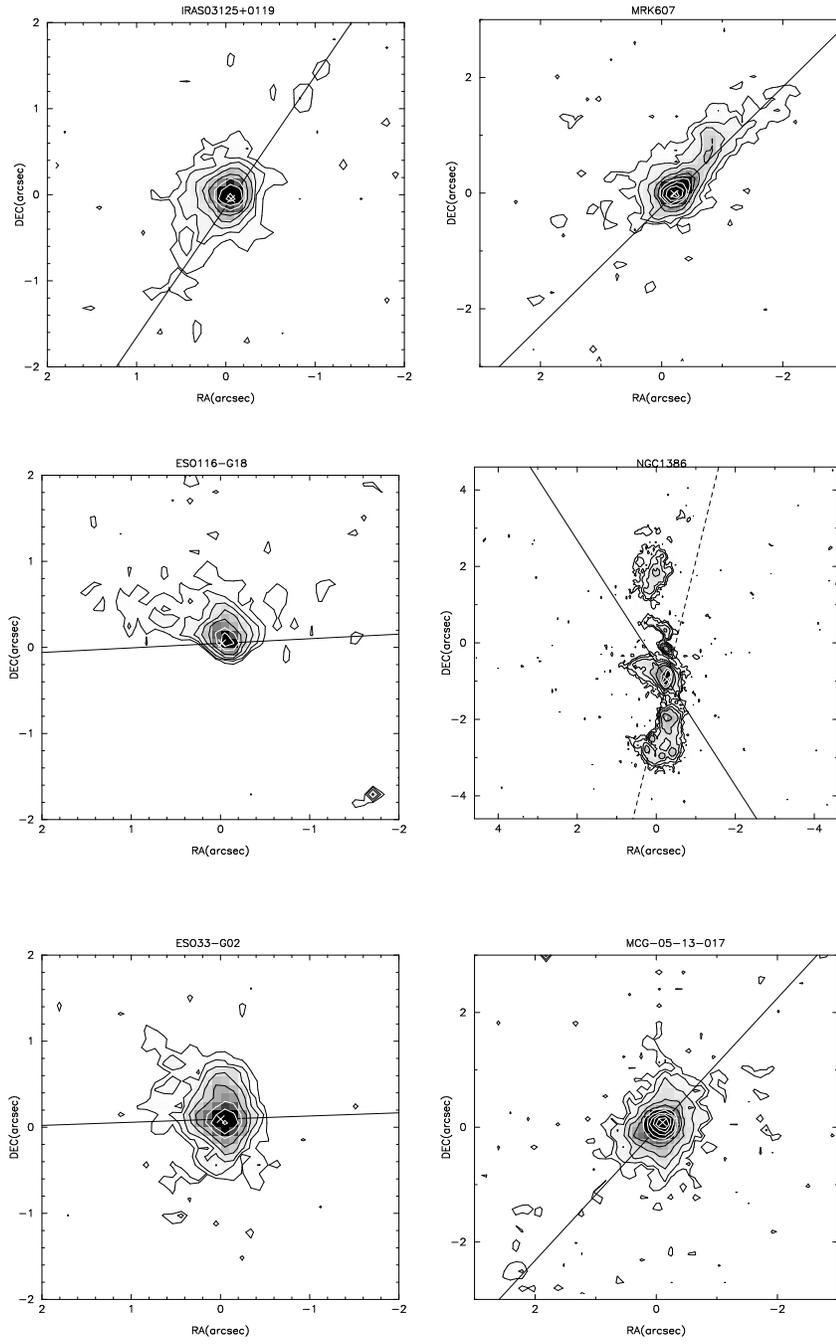}
\vskip -2.0cm
\caption{Same as Figure 5 for IRAS\,03125+0119, MRK\,607, ESO\,116-G\,18,
NGC\,1386, ESO\,33-G\,02 and MCG\,-05-13-017.}
\end{figure}

\begin{figure}
\plotone{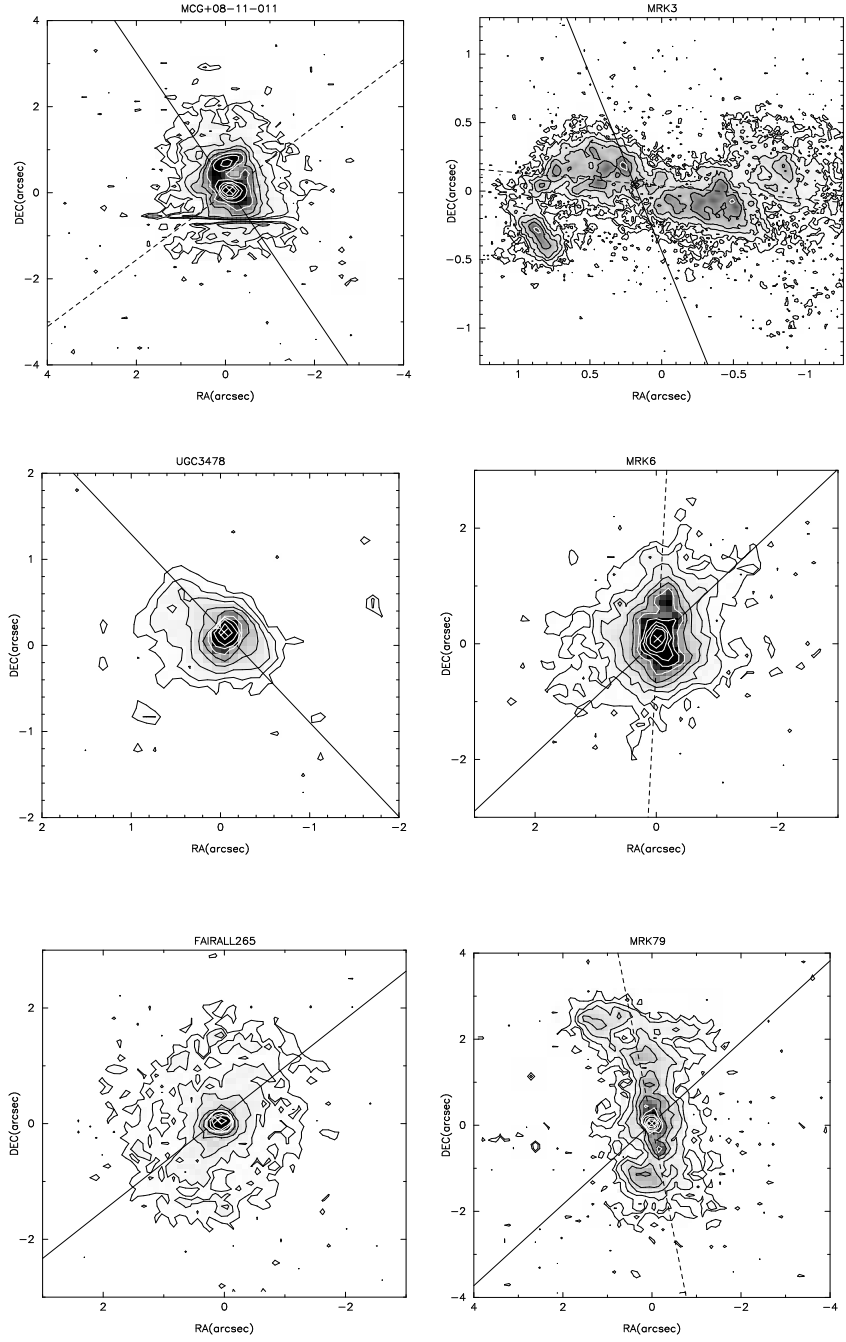}
\vskip -2.0cm
\caption{Same as Figure 5 for MCG\,+08-11-011, MRK\,3, UGC\,3478, MRK\,6,
Fairall\,265 and MRK\,79.}
\end{figure}

\begin{figure}
\plotone{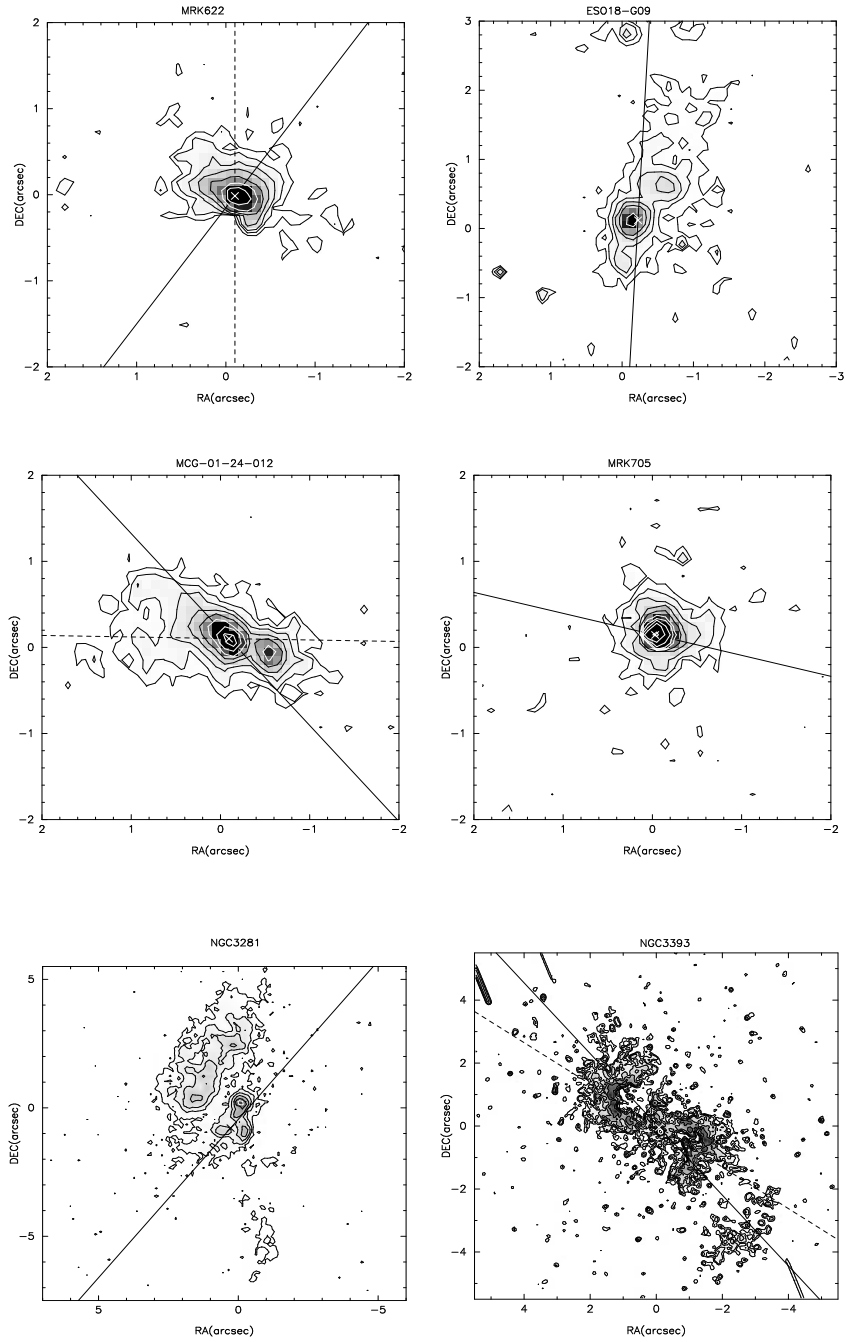}
\vskip -2.0cm
\caption{Same as Figure 5 for MRK\,622, ESO\,18-G\,09, MCG\,-01-24-012,
MRK\,705, NGC\,3281 and NGC\,3393.}
\end{figure}

\vskip -2.0cm
\begin{figure}
\epsscale{0.8}
\plotone{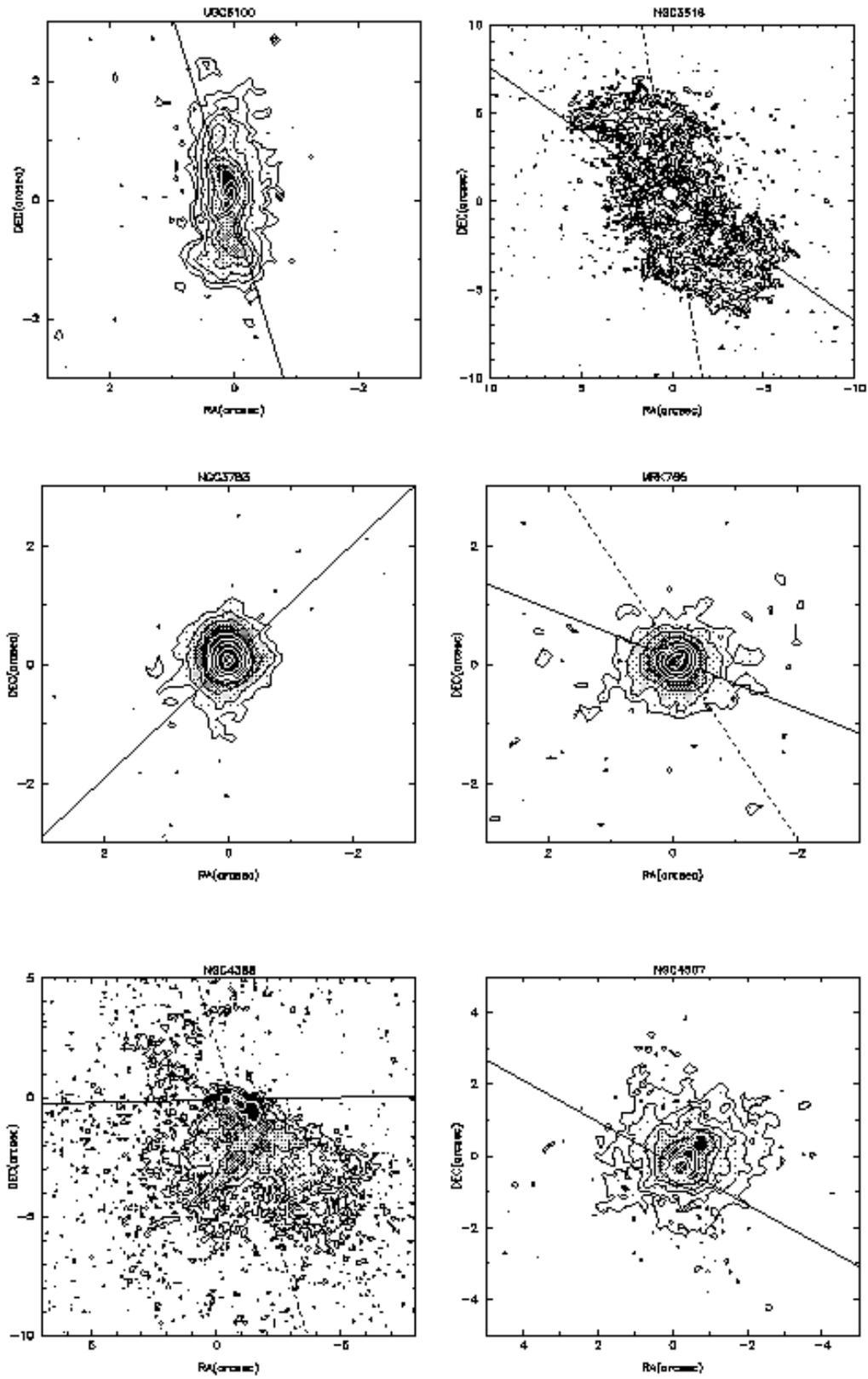}
\vskip -0.05cm
\caption{Same as Figure 5 for UGC\,6100, NGC\,3516, NGC\,3783, MRK\,766,
NGC\,4388 and NGC\,4507.}
\end{figure}

\begin{figure}
\plotone{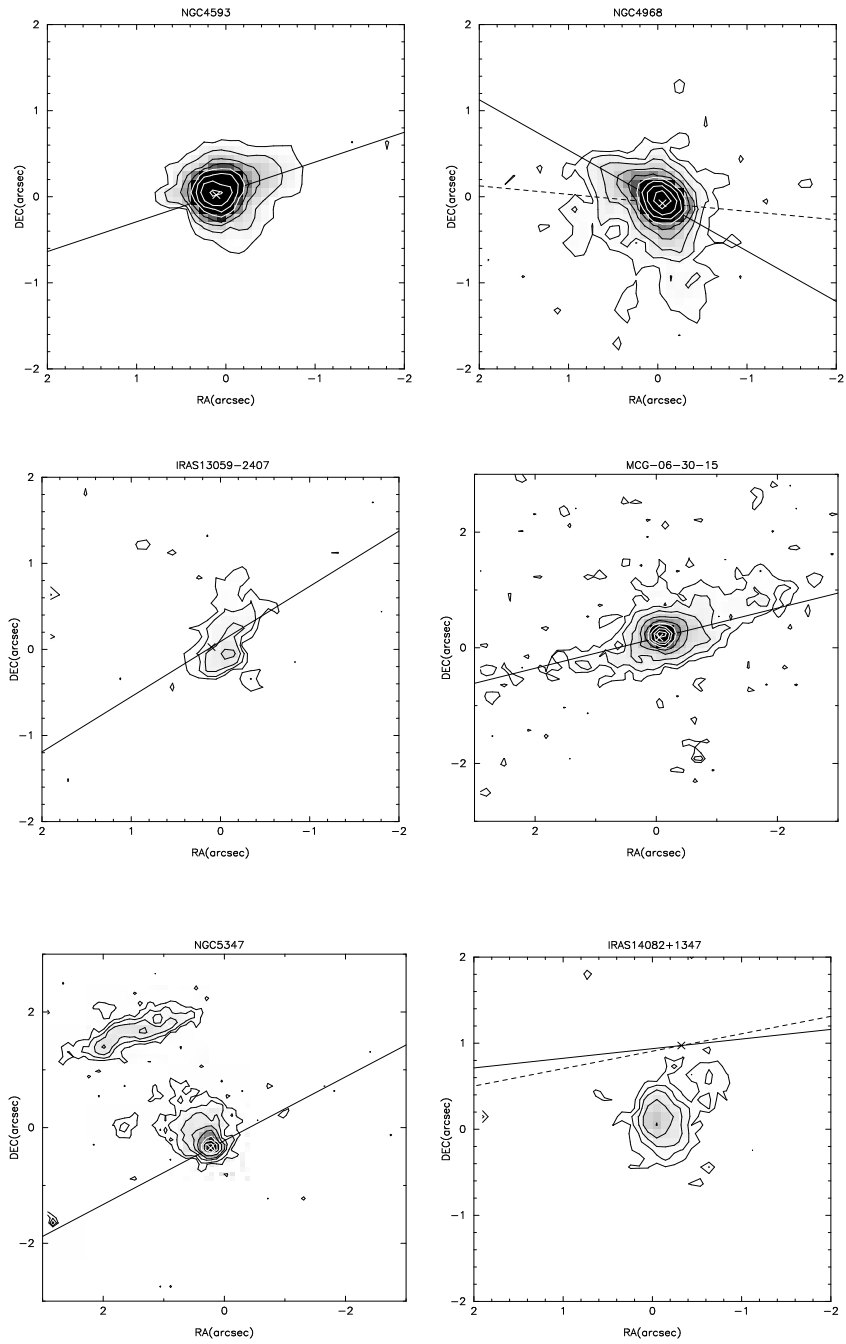}
\vskip -2.0cm
\caption{Same as Figure 5 for NGC\,4593, NGC\,4968, IRAS\,13059-2407,
MCG\,-6-30-15, NGC\,5347, IRAS\,14082+1347.}
\end{figure}

\begin{figure}
\plotone{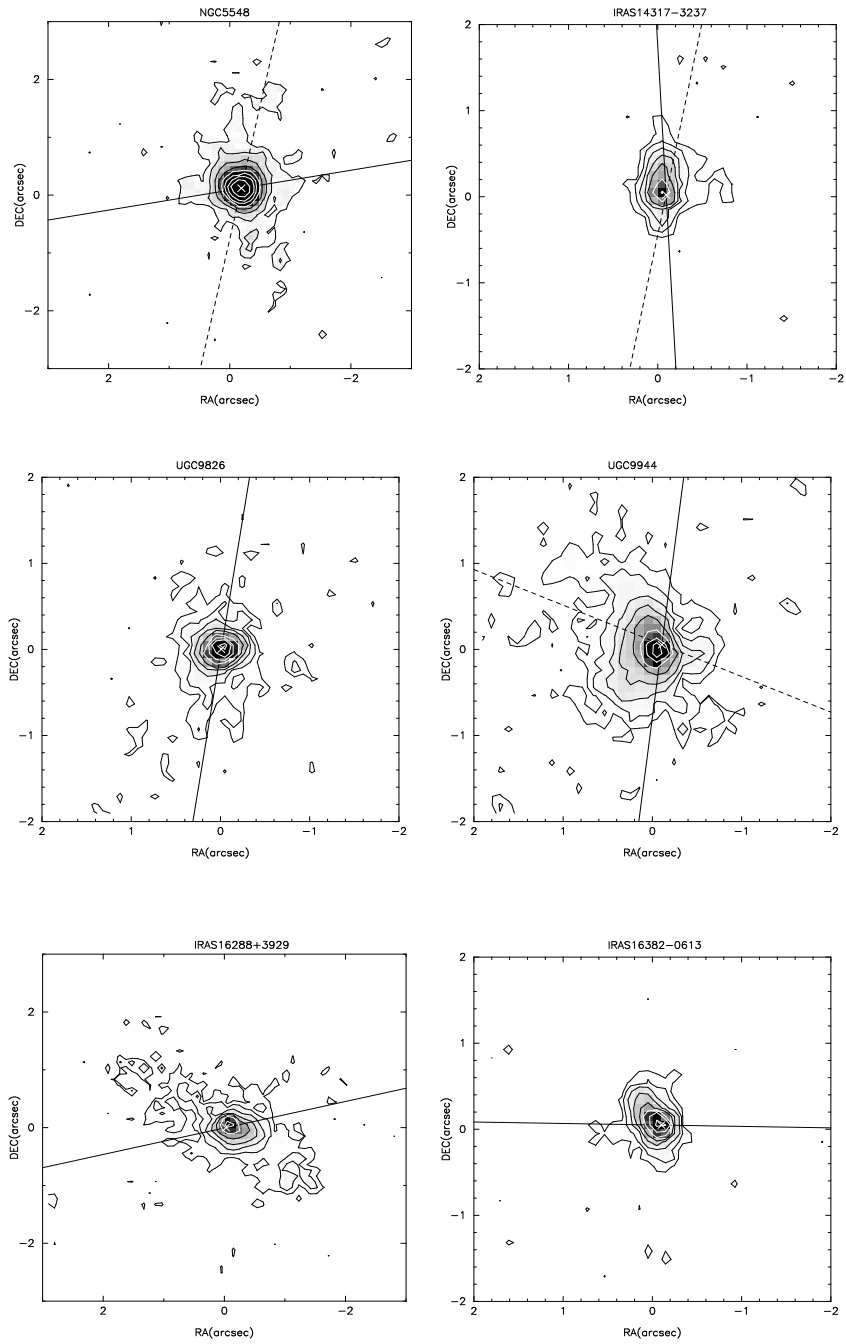}
\vskip -2.0cm
\caption{Same as Figure 5 for NGC\,5548, IRAS\,14317-3237, UGC\,9826, UGC\,9944,
IRAS\,16288+3929, IRAS\,16382-0613.}
\end{figure}

\begin{figure}
\plotone{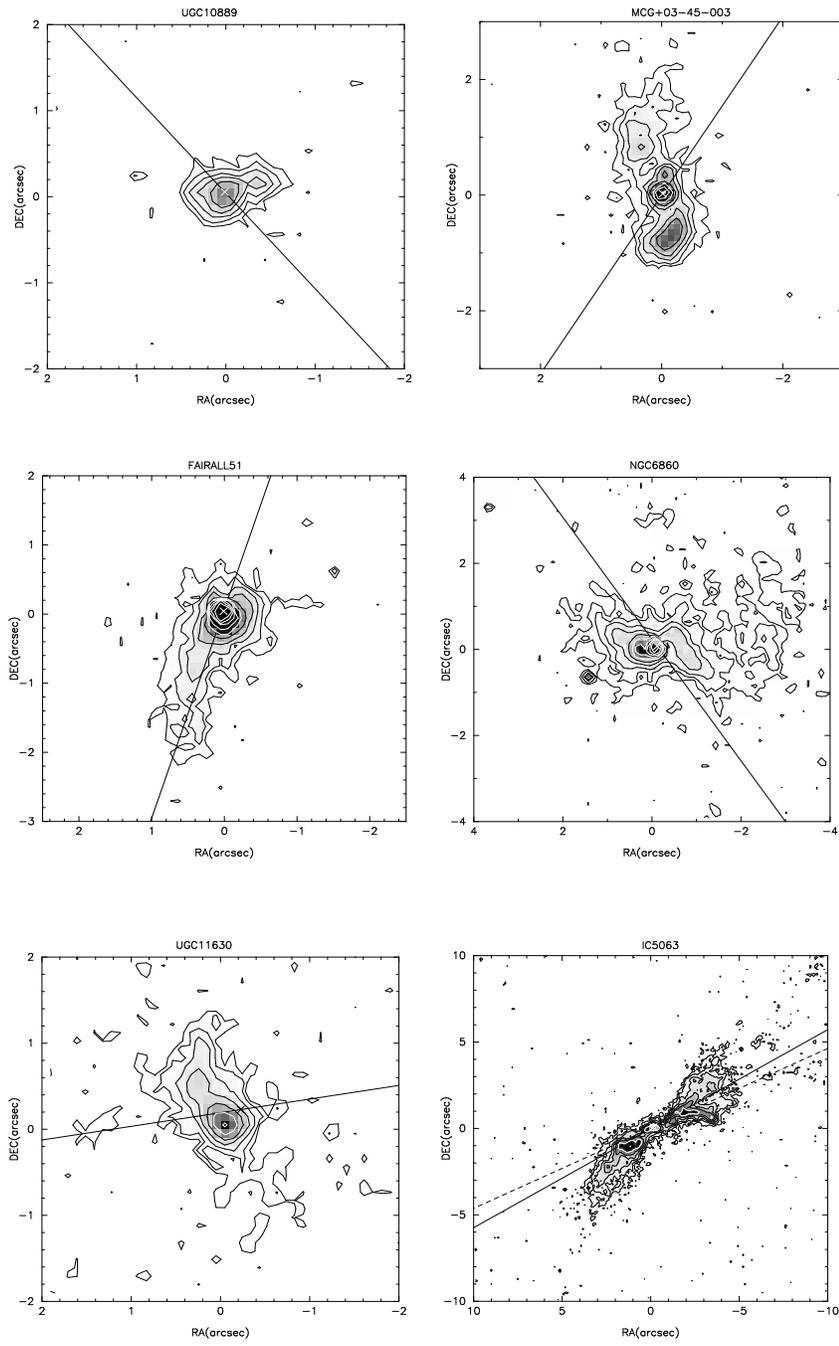}
\vskip -2.0cm
\caption{Same as Figure 5 for UGC\,10889, MCG\,+03-34-003, Fairall\,51,
NGC\,6860, UGC\,11630, IC\,5063.}
\end{figure}

\begin{figure}
\plotone{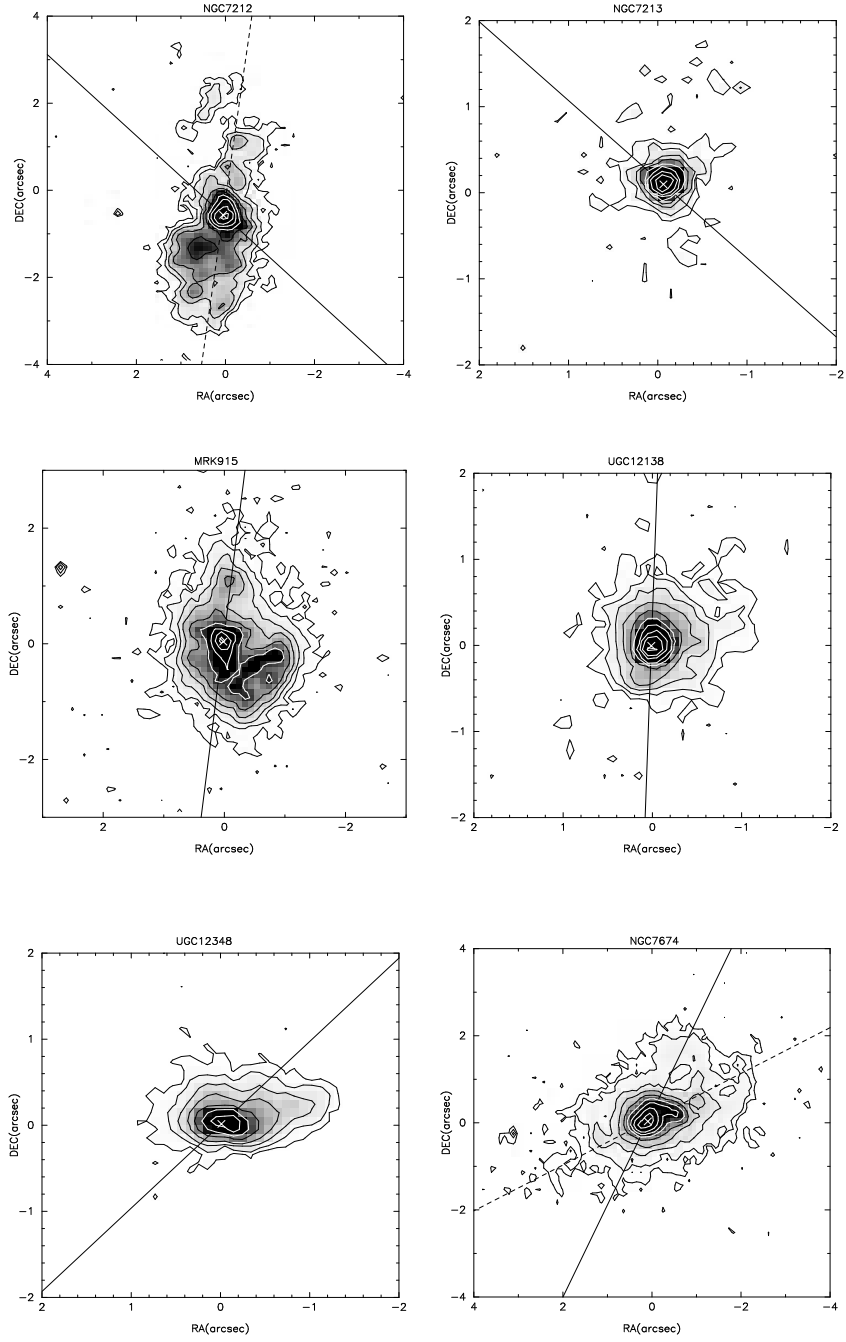}
\vskip -2.0cm
\caption{Same as Figure 5 for NGC\,7212, NGC\,7213, MRK\,915, UGC\,12138,
UGC\,12348, NGC\,7674.}
\end{figure}

\clearpage
\begin{deluxetable}{lrrrlll}
\tabletypesize{\scriptsize}
\tablewidth{0pc}
\tablecaption{Observations Summary}
\tablehead{\colhead{Name}&
\colhead{RA}&
\colhead{DEC}&
\colhead{Exposure}&
\colhead{Dataset Name}&
\colhead{Proposal}&
\colhead{Comments}\\
\colhead{}&
\colhead{(J2000)}&
\colhead{(J2000)}&
\colhead{(seconds)}&
\colhead{}&
\colhead{}&
\colhead{}\\
\colhead{(1)}&
\colhead{(2)}&
\colhead{(3)}&
\colhead{(4)}&
\colhead{(5)}&
\colhead{(6)}&
\colhead{(7)}}
\startdata
MRK\,348          &00 48 47.1 & +31 57 25 &   600 & U2XI0601T,U2XI0602T&6332&a\\
                  &           &           &   280 & U2XI0603T,U2XI0604T&&\\
MRK\,573          &01 43 57.8 & +02 21 00 &   600 & U2XI0701T,U2XI0702T&6332&a\\                  &           &           &   280 & U2XI0703T,U2XI0704T&&\\
IRAS\,01475-0740  &01 50 02.7 & -07 25 48 &   800 & U67L0301R,U67L0302R&8598&\\
                  &           &           &    80 & U67L0303R&&\\
ESO\,153-G\,20    &02 06 03.6 & -55 11 35 &   800 & U67L0401R,U67L0402R&8598&\\
                  &           &           &    80 & U67L0403R&&\\
MRK\,590          &02 14 33.6 & -00 46 00 &   800 & U67L5801R,U67L5802R&8598&\\
                  &           &           &    80 & U67L5803R&&\\
MRK\,1040         &02 28 14.5 & +31 18 42 &   800 & U67L0501R,U67L0502R&8598&\\
                  &           &           &    80 & U67L0503R&&\\
UGC\,2024         &02 33 01.2 & +00 25 15 &  1000 & U67L0701R,U67L0702R&8598&\\
                  &           &           &    80 & U67L0703R&&\\
NGC\,1068         &02 42 40.7 & -00 00 48 &   900 & U2M30103T,U2M30104T&5754&b\\                  &           &           &   440 & U2M30101T,U2M30102T&&\\
MRK\,1058         &02 49 51.8 & +34 59 17 &   800 & U67L5901R,U67L5902R&8598&\\
                  &           &           &    80 & U67L5903R&&\\
MCG\,-02-08-039   &03 00 29.8 & -11 24 59 &   600 & U67L0801R,U67L0802R&8598&c\\                  &           &           &    80 & U67L0803R&&\\
UGC\,2514         &03 03 48.5 & -01 06 13 &  1000 & U67L0901R,U67L0902M&8598&\\
                  &           &           &    80 & U67L0903R&&\\
IRAS\,03106-0254  &03 13 08.3 & -02 43 19 &  1000 & U67L1001R,U67L1002R&8598&\\
                  &           &           &    80 & U67L1003R&&\\
IRAS\,03125+0119  &03 15 05.3 & +01 30 30 &   800 & U67L1101R,U67L1102R&8598&\\
                  &           &           &    80 & U67L1103R&&\\
MRK\,607          &03 24 48.7 & -03 02 33 &   400 & U3A00101T,U3A00102T&6419&\\
                  &           &           &    80 & U3A00103T,U3A00104T&&\\
ESO\,116-G\,18    &03 24 53.1 & -60 44 20 &   800 & U67L1201R,U67L1202R&8598&\\
                  &           &           &    80 & U67L1203R&&\\
NGC\,1386         &03 36 45.4 & -35 59 57 &   800 & U3A00201M,U3A00202M&6419&b\\                  &           &           &    80 & U3A00203M,U3A00204M&&\\
ESO\,33-G\,02     &04 55 59.6 & -75 32 27 &   800 & U67L1501R,U67L1502R&8598&\\
                  &           &           &    80 & U67L1503R&&\\
MCG\,-05-13-017   &05 19 35.5 & -32 39 30 &   600 & U67L1601R,U67L1602R&8598&\\
                  &           &           &    80 & U67L1603R&&\\
MCG\,+08-11-011   &05 54 53.6 & +46 26 22 &   600 & U67L6001R,U67L6002R&8598&\\
                  &           &           &    80 & U67L6003R&&\\
MRK\,3            &06 15 36.3 & +71 02 15 &   750 & X2580103T&5140&d\\
                  &           &           &  1196 & X2580104T&&\\
UGC\,3478         &06 32 47.3 & +63 40 25 &   800 & U67L1701R,U67L1702R&8598&\\
                  &           &           &    80 & U67L1703R&&\\
MRK\,6            &06 52 12.3 & +74 25 37 &   600 & U67L1801R,U67L1802R&8598&\\
                  &           &           &    80 & U67L1803R&&\\
FAIRALL\,265      &06 56 29.7 & -65 33 43 &  1000 & U67L1901R,U67L1902R&8598&c\\                  &           &           &    80 & U67L1903R&&\\
MRK\,79           &07 42 32.8 & +49 48 35 &   600 & U67L2001R,U67L2002R&8598&\\
                  &           &           &    80 & U67L2003R&&\\
MRK\,622          &08 07 41.0 & +39 00 15 &   800 & U67L2301R,U67L2302R&8598&\\
                  &           &           &    80 & U67L2303R&&\\
ESO\,18-G\,09     &08 24 07.4 & -77 46 53 &  1000 & U67L2401R,U67L2402R&8598&\\
                  &           &           &    80 & U67L2403R&&\\
MCG\,-01-24-012   &09 20 51.7 & -08 04 47 &   800 & U67L2501R,U67L2502R&8598&\\
                  &           &           &    80 & U67L2503R&&\\
MRK\,705          &09 26 03.3 & +12 44 04 &   800 & U67L6101N,U67L6102R&8598&c\\                  &           &           &    80 & U67L6103R&&\\
NGC\,3281         &10 31 52.1 & -34 51 13 &   800 & U67L6201R,U67L6202R&8598&\\
                  &           &           &    80 & U67L6203R&&\\
NGC\,3393         &10 48 24.0 & -25 09 40 & 10000 & W1DB0405T,W1DB0406T,W1DB0407T&3982&e\\
                  &           &           &       & W1DB0408T,W1DB0409T&&\\
                  &           &           &   600 & W1DB0401T&&\\
UGC\,6100         &11 01 34.0 & +45 39 14 &  1000 & U67L6301R,U67L6302R&8598&c\\                  &           &           &    80 & U67L6303R&&\\
NGC\,3516         &11 06 47.5 & +72 34 07 &  1600 & U3A00601T,U3A00602T&6419&\\
                  &           &           &   140 & U3A00603T,U3A00604T&&\\
IRAS\,11215-2806  &11 24 02.6 & -28 23 15 &     1 & U67L2601R,U67L2602R&8598&f\\                  &           &           &    80 & U67L2603R&&\\
NGC\,3783         &11 39 01.8 & -37 44 20 &   400 & U5GU0105R,U5GU0106R&8240&g\\                  &           &           &       & U5GU0107M,U5GU0108R&&\\
                  &           &           &    16 & U5GU0103R,U5GU0104R&&\\
MRK\,766          &12 18 26.5 & +29 48 46 &   480 & U5GU0205R,U5GU0206R&8240&g\\                  &           &           &       & U5GU0207R,U5GU0208R&&\\
                  &           &           &    20 & U5GU0203R,U5GU0204R&&\\
NGC\,4388         &12 25 46.7 & +12 39 41 &   280 & U2XI0201T,U2XI0202T&6332&a\\                  &           &           &   280 & U2XI0203T,U2XI0204T&&\\
NGC\,4507         &12 35 37.0 & -39 54 31 &   520 & U5DF0105R,U5DF0106R&8259&g\\                  &           &           &   200 & U5DF0101R,U5DF0102R&&\\
NGC\,4593         &12 39 39.4 & -05 20 39 &   320 & U5GU0306R,U5GU0307M&8240&g\\                  &           &           &       & U5GU0308R,U5GU0309R&&\\
                  &           &           &   300 & U5GU0305R&&\\
NGC\,4968         &13 07 07.6 & -23 40 43 &   400 & U3A00801P,U3A00802P&6419&\\
                  &           &           &    80 & U3A00803P,U3A00804P&&\\
IRAS\,13059-2407  &13 08 42.1 & -24 23 00 &  1000 & U67L3401R,U67L3402R&8598&\\
                  &           &           &    80 & U67L3403R&&\\
MCG\,-6-30-15     &13 35 53.7 & -34 17 45 &   400 & U3A00901T,U3A00902T&6419&\\
                  &           &           &    80 & U3A00903T,U3A00904T&&\\
NGC\,5347         &13 53 17.8 & +33 29 27 &   800 & U67L3701R,U67L3702R&8598&\\
                  &           &           &    80 & U67L3703R&&\\
IRAS\,14082+1347  &14 10 41.4 & +13 33 29 &  1000 & U67L3801R,U67L3802R&8598&\\
                  &           &           &    80 & U67L3803R&&\\
NGC\,5548         &14 17 59.5 & +25 08 12 &   480 & U5GU0505R,U5GU0506R&8240&g\\                  &           &           &       & U5GU0507R,U5GU0508R&&\\
                  &           &           &    24 & U5GU0503R,U5GU0504R&&\\
IRAS\,14317-3237  &14 34 44.9 & -32 50 28 &  1000 & U67L3901R,U67L3902R&8598&\\
                  &           &           &    80 & U67L3903R&&\\
UGC\,9826         &15 21 33.0 & +39 12 01 &   800 & U67L4101R,U67L4102R&8598&c\\                  &           &           &    80 & U67L4103R&&\\
UGC\,9944         &15 35 47.8 & +73 27 02 &   800 & U67L4201R,U67L4202R&8598&\\
                  &           &           &    80 & U67L4203R&&\\
IRAS\,16288+3929  &16 30 32.6 & +39 23 03 &   800 & U67L4401R,U67L4402R&8598&c\\                  &           &           &    80 & U67L4403R&&\\
IRAS\,16382-0613  &16 40 52.3 & -06 18 52 &  1000 & U67L4501R,U67L4502R&8598&\\
                  &           &           &    80 & U67L4503R&&\\
UGC\,10683\,B     &17 05 00.4 & -01 32 29 &     1 & U67L6501R,U67L6502R&8598&f\\                  &           &           &    80 & U67L6503R&&\\
UGC\,10889        &17 30 20.7 & +59 38 20 &  1000 & U67L4601R,U67L4602R&8598&\\
                  &           &           &    80 & U67L4603R&&\\
MCG\,+03-45-003   &17 35 32.7 & +20 47 48 &   600 & U67L7001R,U67L7002R&8598&\\
                  &           &           &    80 & U67L7003R&&\\
FAIRALL\,49       &18 36 58.1 & -59 24 09 &   600 & U67L4701R,U67L4702R&8598&h\\                  &           &           &    80 & U67L4703R&&\\
FAIRALL\,51       &18 44 54.3 & -62 21 49 &   600 & U67L4801R,U67L4802R&8598&\\
                  &           &           &    80 & U67L4803R&&\\
NGC\,6860         &20 08 46.1 & -61 05 56 &  1000 & U67L4901R,U67L4902R&8598&\\
                  &           &           &    80 & U67L4903R&&\\
UGC\,11630        &20 47 33.5 & +00 24 42 &  1000 & U67L5101R,U67L5102R&8598&\\
                  &           &           &    80 & U67L5103M&&\\
IC\,5063          &20 52 02.0 & -57 04 09 &   600 & U67L5201M,U67L5202R&8598&\\
                  &           &           &    80 & U67L5203R&&\\
NGC\,7212         &22 07 02.0 & +10 14 00 &   600 & U2XI0401T,U2XI0402T&6332&a\\                  &           &           &   280 & U2XI0403T,U2XI0404T&&\\
NGC\,7213         &22 09 16.2 & -47 10 00 &   600 & U67L5301R,U67L5302R&8598&\\
                  &           &           &    80 & U67L5303R&&\\
MRK\,915          &22 36 46.5 & -12 32 45 &   600 & U67L5401R,U67L5402R&8598&\\
                  &           &           &    80 & U67L5403R&&\\
UGC\,12138        &22 40 17.0 & +08 03 14 &   600 & U67L5501R,U67L5502R&8598&\\
                  &           &           &    80 & U67L5503R&&\\
UGC\,12348        &23 05 19.4 & +00 11 28 &   600 & U67L5701R,U67L5702R&8598&\\
                  &           &           &    80 & U67L5703R&&\\
NGC\,7674         &23 27 56.7 & +08 46 45 &   520 & U5DF0206R,U5DF0205R&8259&i\\                  &           &           &   200 & U5DF0201R,U5DF0202R&&\\
\enddata
\tablenotetext{a}{On-band and off-band images were obtained on different
WF chips.}
\tablenotetext{b}{On-band image obtained using the filter F502N, with the
galaxy centered on the PC chip.}
\tablenotetext{c}{Used a redshift slightly smaller than the observed
one to force the galaxy to fall on the WF camera instead of the
PC camera.}
\tablenotetext{d}{Images obtained with FOC camera.}
\tablenotetext{e}{Images obtained with the WF/PC1 in the PC mode, with the
filter centered at [OIII]$\lambda$4959\AA.}
\tablenotetext{f}{Observation failed becasue of guide star acquisition
problems.}
\tablenotetext{g}{On-band observed with the WF and the off-band with the
PC camera.}
\tablenotetext{h}{Galaxy located on a bad column on the CCD. Data lost.}
\tablenotetext{i}{Observations taken with the PC camera.}
\end{deluxetable}

\clearpage

\begin{deluxetable}{lllrrrrrrr}
\tabletypesize{\scriptsize}
\tablewidth{0pc}
\tablecaption{Sample Characteristics}
\tablehead{\colhead{Name}&
\colhead{Alternative Name}&
\colhead{Type}&
\colhead{V$_{Rad}$$^a$}&
\colhead{$3\sigma^b$}&
\colhead{F([OIII])$_{int}$$^c$}&
\colhead{F([OIII])$_{nuc}$$^c$}&
\colhead{F([OIII])$_{lit}$$^c$}&
\colhead{Log L([OIII])$^d$}&
\colhead{Ref.$^e$}\\
\colhead{(1)}&
\colhead{(2)}&
\colhead{(3)}&
\colhead{(4)}&
\colhead{(5)}&
\colhead{(6)}&
\colhead{(7)}&
\colhead{(8)}&
\colhead{(9)}&
\colhead{(10)}}
\startdata
MRK\,348        &NGC\,262         &2 &4540 &10.0 &41.2  &30.33 &36.0 &41.26&1\\
MRK\,573        &UGC\,1214        &2 &5174 &6.00 &159.0 &31.15 &93.0 &41.96&1\\
IRAS\,01475-0740&                 &2 &5306 &3.91 &5.44  &5.12  &5.3 &40.51&1\\
ESO\,153-G\,20  &IRAS\,02043-5525 &2 &5917 &3.43 &6.34  &1.81  &4.2 &40.68&1\\
MRK\,590        &NGC\,863         &1 &7910 &3.51 &13.6  &11.29 &5.3 &41.26&2\\
MRK\,1040       &NGC\,931         &1 &4927 &4.10 &10.9  &8.98  &7.5 &40.75&1\\
UGC\,2024       &                 &2 &6714 &3.00 &5.50  &2.24  &2.5 &40.72&1\\
NGC\,1068       &                 &2 &1136 &2.75 &1220.0&808.4 &484.0&41.53&1\\
MRK\,1058       &IRAS\,02467+3446 &2 &5138 &3.97 &5.49  &2.47  &5.6 &40.49&2\\
MCG\,-02-08-039 &IRAS\,02580-1136 &2 &8874 &4.60 &19.5  &13.40 &18.0 &41.51&1\\
UGC\,2514       &NGC\,1194        &1 &3957 &3.36 &1.20  &0.46  &1.4 &39.60&1\\
IRAS\,03106-0254&                 &2 &8154 &3.05 &3.58  &1.41  &2.3 &40.71&1\\
IRAS\,03125+0119&                 &2 &7200 &4.00 &4.97  &3.14  &6.1 &40.74&1\\
MRK\,607        &NGC\,1320        &2 &2716 &5.96 &17.6  &14.53 &12.0 &40.44&1\\
ESO\,116-G\,18  &IRAS\,03238-6054 &2 &5546 &3.78 &2.97  &1.94  &5.2 &40.29&1\\
NGC\,1386       &                 &2 & 868 &4.38 &51.4  &38.62 &80.0 &39.92&2\\
ESO\,33-G\,02   &IRAS\,04575-7537 &2 &5426 &3.85 &6.21  &4.13  &5.7 &40.59&1\\
MCG\,-05-13-017 &IRAS\,05177-3242 &1 &3790 &4.47 &30.7  &26.51 &34.0 &40.97&1\\
MCG\,+08-11-011 &UGC\,3374        &1 &6141 &3.26 &64.3  &34.23 &71.0 &41.71&2\\
MRK\,3          &                 &2 &4050 &4.31 &205.0 &57.10 &107.0&41.86&1\\
UGC\,3478       &                 &1 &3828 &3.53 &8.84  &7.63  &5.8 &40.44&1\\
MRK\,6          &IC\,450          &1 &5537 &5.48 &79.6  &49.87 &70.0 &41.72&1\\
Fairall\,265    &IRAS\,06563-6529 &1 &8844 &4.51 &18.3  &9.17  &2.8 &41.48&1\\
MRK\,79         &UGC\,3973        &1 &6652 &3.94 &40.6  &19.16 &37.0 &41.58&2\\
MRK\,622        &UGC\,4229        &2 &6964 &3.95 &5.08  &2.85  &4.0 &40.72&2\\
ESO\,18-G\,09   &IRAS\,08255-7737 &2 &5341 &3.18 &4.21  &1.91  &3.2 &40.41&1\\
MCG\,-01-24-012 &IRAS\,09182-0750 &2 &5892 &3.42 &7.16  &3.77  &8.9 &40.72&1\\
MRK\,705        &UGC\,5025        &1 &8658 &4.06 &12.4  &9.69  &9.0 &41.30&2\\
NGC\,3281       &                 &2 &3200 &4.10 &25.0  &1.61  &5.5 &40.74&2\\
NGC\,3393       &                 &2 &4107 &2.38 &105.0 &5.97  &99.0 &41.58&1\\
UGC\,6100       &                 &2 &8778 &3.59 &20.5  &7.13  &28.0 &41.53&3\\
NGC\,3516       &                 &1 &2649 &1.94 &69.3  &23.47 &35.0 &41.02&1\\
NGC\,3783       &                 &1 &2550 &16.5 &86.3  &81.63 &76.0 &41.08&1\\
MRK\,766        &NGC\,4253        &1 &3876 &10.5 &39.5  &31.55 &45.0 &41.10&1\\
NGC\,4388       &                 &2 &2524 &7.87 &134.0 &21.55 &66.0 &41.26&1\\
NGC\,4507       &                 &2 &3538 &4.26 &110.0 &78.29 &83.0 &41.47&1\\
NGC\,4593       &                 &1 &2698 &13.6 &22.9  &21.80 &13.0 &40.55&1\\
NGC\,4968       &MCG\,-04-31-030  &2 &2957 &5.77 &20.4  &18.55 &18.0 &40.58&1\\
IRAS\,13059-2407&                 &2 &4175 &2.97 &0.63  &0.37  &1.2 &39.37&1\\
MCG\,-6-30-15   &                 &1 &2323 &4.00 &14.8  &12.39 &7.5 &40.23&1\\
NGC\,5347       &                 &2 &2335 &3.71 &7.92  &5.27  &4.5 &39.96&1\\
IRAS\,14082+1347&                 &2 &4836 &3.23 &1.02  &0.02  &2.8 &39.71&1\\
NGC\,5548       &                 &1 &5149 &10.5 &41.3  &32.86 &73.0 &41.37&1\\
IRAS\,14317-3237&                 &2 &7615 &3.00 &1.75  &0.92  &2.1 &40.34&1\\
UGC\,9826       &                 &1 &8754 &3.18 &5.12  &3.47  &5.2 &40.92&1\\
UGC\,9944       &                 &2 &7354 &3.44 &6.87  &2.78  &4.7 &40.90&1\\
IRAS\,16288+3929&                 &2 &9091 &3.62 &4.83  &1.33  &2.2 &40.93&1\\
IRAS\,16382-0613&                 &2 &8317 &3.08 &2.87  &1.94  &2.8 &40.63&1\\
UGC\,10889      &NGC\,6393        &2 &8424 &2.82 &1.38  &0.65  &2.2 &40.32&1\\
MCG\,+03-45-003 &IRAS\,17334+2049 &2 &7292 &4.70 &9.58  &2.91  &26.0 &41.04&1\\
Fairall\,51     &IRAS\,18401-6225 &1 &4255 &4.38 &27.2  &24.34 &19.0 &41.02&1\\
NGC\,6860       &ESO\,143-G\,09   &1 &4462 &2.20 &18.1  &10.75 &2.5 &40.89&1\\
UGC\,11630      &NGC\,6967        &2 &3657 &3.20 &3.18  &1.68  &2.6 &39.96&1\\
IC\,5063        &PKS\,2048-57     &2 &3402 &4.04 &77.2  &9.95  &56.0 &41.28&1\\
NGC\,7212       &                 &2 &7984 &13.8 &87.5  &21.28 &70.0 &42.08&2\\
NGC\,7213       &                 &1 &1792 &10.0 &13.6  &13.09 &13.0 &39.97&1\\
MRK\,915        &IRAS\,22340-1248 &1 &7230 &5.17 &47.3  &9.53 &31.0 &41.72&1\\
UGC\,12138      &                 &1 &7375 &5.37 &21.9  &14.68 &14.0 &41.40&1\\
UGC\,12348      &                 &2 &7585 &6.47 &10.1  &4.22  &13.0 &41.09&1\\
NGC\,7674       &MRK\,533         &2 &8713 &3.22 &52.1  &12.03 &52.1 &41.93&1\\
\enddata
\tablenotetext{a}{The units of Column (4) are km~s$^{-1}$.}
\tablenotetext{b}{The units of Column (5), the 3$\sigma$ surface brightness
level of the images, are 10$^{-17}$~erg~cm$^{-2}$~s$^{-1}$~pix$^{-1}$.}
\tablenotetext{c}{The Units of Column (6), the integrated [OIII] flux,
Column (7), the nuclear [OIII] flux, and Column (8), the [OIII] flux from
the literature, are 10$^{-14}$~erg~cm$^{-2}$~s$^{-1}$.}
\tablenotetext{d}{The units of Column (9), the integrated [OIII] luminosity
are erg~s$^{-1}$.}
\tablenotetext{e}{References for the [OIII] data from the literature:
1-) de Grijp et al. (1992); 2-) Whittle (1992); 3-) Cruz-Gonzalez et al. 1994.}
\end{deluxetable}

\begin{deluxetable}{lrrrrrr}
\tabletypesize{\footnotesize}
\tablewidth{0pc}
\tablecaption{NLR Sizes}
\tablehead{\colhead{Name}&
\colhead{R$_e$}&
\colhead{R$_{Maj}$}&
\colhead{R$_{Min}$}&
\colhead{P.A.}&
\colhead{Offsets}&
\colhead{Figure}\\
\colhead{}&
\colhead{(pc)}&
\colhead{(pc)}&
\colhead{(pc)}&
\colhead{($^{\circ}$)}&
\colhead{}&
\colhead{}\\
\colhead{(1)}&
\colhead{(2)}&
\colhead{(3)}&
\colhead{(4)}&
\colhead{(5)}&
\colhead{(6)}&
\colhead{(7)}}
\startdata
MRK\,348           &  56&420   &255   &185 & 0.16&5\\
MRK\,573           & 505&1490  &625   &120 & 0.13&5\\
IRAS\,01475-0740   &  33&145   &130   &--- & 0.00&5\\
ESO\,153-G\,20     & 163&505   &355   &-10 (-100) & 0.43&5\\
MRK\,590           &  49&385   &280   &-5  & 0.30&5\\
MRK\,1040          &  40&240   &175   &45 (130) & 0.00&5\\
UGC\,2024          & 122&565   &315   &-35 & 0.10&6\\
NGC\,1068          &  61&375   &215   &35  & 0.23&6\\
MRK\,1058          & 123&380   &235   &205 & 0.56&6\\
MCG\,-02-08-039    &  75&630   &530   &--- & 0.08&6\\
UGC\,2514          & 127&175   &145   &110 & 0.71&6\\
IRAS\,03106-0254   & 119&395   &235   &30  & 0.33&6\\
IRAS\,03125+0119   &  74&350   &235   &145 & 0.33&7\\
MRK\,607           &  31&330   &120   &-45 & 0.44&7\\
ESO\,116-G\,18     &  66&305   &180   &75  & 0.56&7\\
NGC\,1386          &  46&165   &45    &0   & 0.20&7\\
ESO\,33-G\,02      &  70&255   &175   &-5  & 0.09&7\\
MCG\,-05-13-017    &  32&250   &190   &-40 & 0.11&7\\
MCG\,+08-11-011    &  87&725   &605   &20  & 0.24&8\\
MRK\,3             & 147&290   &130   &90  & 0.33&8\\
UGC\,3478          &  32&205   &145   &55  & 0.09&8\\
MRK\,6             &  68&575   &390   &-10 & 0.08&8\\
Fairall\,265       & 100&985   &830   &-20 & 0.04&8\\
MRK\,79            & 129&990   &460   &15  & 0.29&8\\
MRK\,622           &  88&290   &215   &55  & 0.23&9\\
ESO\,18-G\,09      & 135&520   &190   &-30 & 0.37&9\\
MCG\,-01-24-012    &  93&440   &220   &75  & 0.09&9\\
MRK\,705           &  62&350   &280   &--- & 0.22&9\\
NGC\,3281          & 430&630   &345   &5 (50)  & 0.68&9\\
NGC\,3393          & 446&705   &370   &65  & 0.01&9\\
UGC\,6100          & 145&995   &370   &0   & 0.12&10\\
NGC\,3516          & 361&1165  &590   &20  & 0.10&10\\
NGC\,3783          &  31&175   &140   &-10 & 0.00&10\\
MRK\,766           &  50&255   &215   &65  & 0.02&10\\
NGC\,4388          & 385&635   &500   &90 (205) & 0.80&10\\
NGC\,4507          &  50&400   &375   &-35 & 0.30&10\\
NGC\,4593          &  31&150   &120   &100 & 0.19&11\\
NGC\,4968          &  31&210   &125   &40  & 0.14&11\\
IRAS\,13059-2407   &  84&180   &70    &-20 & 0.47&11\\
MCG\,-6-30-15      &  27&295   &115   &-65 & 0.23&11\\
NGC\,5347          &  39&235   &185   &30  & 0.78&11\\
IRAS\,14082+1347   & 153&155   &115   &165 & 1.00&11\\
NGC\,5548          &  57&450   &350   &5   & 0.09&12\\
IRAS\,14317-3237   &  94&345   &185   &-5  & 0.29&12\\
UGC\,9826          &  65&435   &255   &165 & 0.26&12\\
UGC\,9944          & 128&475   &355   &0 (90)  & 0.07&12\\
IRAS\,16288+3929   & 191&795   &470   &65  & 0.17&12\\
IRAS\,16382-0613   &  70&325   &240   &15  & 0.15&12\\
UGC\,10889         & 109&340   &220   &-85 & 0.18&13\\
MCG\,+03-45-003    & 269&705   &300   &15  & 0.25&13\\
Fairall\,51        &  29&635   &170   &160 (210) & 0.59&13\\
NGC\,6860          &  62&650   &345   &85  & 0.19&13\\
UGC\,11630         &  30&190   &100   &30  & 0.16&13\\
IC\,5063           & 462&1320  &330   &-65 & 0.04&13\\
NGC\,7212          & 225&1240  &540   &170 & 0.17&14\\
NGC\,7213          &  14&65    &65    &--  & 0.13&14\\
MRK\,915           & 280&955   &610   &5   & 0.08&14\\
UGC\,12138         &  67&395   &345   &130 & 0.21&14\\
UGC\,12348         & 123&540   &280   &100 & 0.23&14\\
NGC\,7674          &  86&1240  &760   &120 & 0.25&14\\
\enddata
\end{deluxetable}

\end{document}